\documentclass[12pt]{article}
\usepackage{epsf}
\title{ {\bf The effect of localized new Higgs doublet on the radiative lepton flavor
violating decays in the Randall Sundrum background}}
\author{\vspace{1cm}\\
        {\bf E. O. Iltan}
        \thanks{E-mail address:
        eiltan@newton.physics.metu.edu.tr}
 \\
        Middle East Technical University,
        Northern Cyprus Campus, Guzelyurt, \\ Mersin 10, TURKEY\\}

\date{}

\begin{document}
\setlength{\baselineskip}{24pt}
\maketitle
\setlength{\baselineskip}{7mm}
\begin{abstract}
We study  the radiative lepton flavor violating $l_1\rightarrow
l_2\gamma$ decays in the two Higgs doublet model with localized
new Higgs doublet in the Randall Sundrum background. We estimate
the contributions of the KK modes of new Higgs bosons and left
(right) handed charged lepton doublets (singlets) on the branching
ratios of the decays considered. We observe that there is an
enhancement in the branching ratios with the addition of new Higgs
boson and lepton KK modes.
\end{abstract}
\thispagestyle{empty}
\newpage
\setcounter{page}{1}
%
\section{Introduction}
The processes with flavor violation (FV) are worthwhile to study
since they exist at least in the one loop level in the standard
model (SM) and, therefore, they are rich from the theoretical
point of view. The lepton flavor violating (LFV) interactions are
among the most exciting candidates of these processes, since they
are clean in the sense that they are free from strong
interactions. Furthermore, the small numerical values of branching
ratios (BRs) of LFV decays stimulate one to search beyond and to
study the more fundamental models in order to enhance these
numerical values to reach the current experimental upper limits.
Among the LFV decays the radiative $l_1\rightarrow l_2\gamma$
processes reach great interest and their current experimental
upper limits of the BRs are: BR $(\mu\rightarrow
e\gamma)=1.2\times 10^{-11}$ \cite{Brooks}, BR $(\tau\rightarrow
e\gamma)=3.9\times 10^{-7}$ \cite{Hayasaka} and BR
$(\tau\rightarrow \mu\gamma)=1.1\times 10^{-6}\, (9.0\times
10^{-8}\,\mathbf{;}\,6.8\times 10^{-8}\mathbf{,}\,\, 90\% CL)$
\cite{Ahmed} (\cite{Roney}; \cite{Aubert}), respectively.
Furthermore, in order to search the $\mu\rightarrow e \gamma$
decay, a new experiment, aiming to reach a sensitivity of BR$\sim
10^{-14}$, at PSI has been described \cite{Nicolo}. At present,
this experiment (PSI-R-99-05 Experiment) is still running in the
MEG \cite{Yamada}.

The theoretical values of the BRs of the radiative LFV decays in
the framework of the SM are negligible compared to the
experimental upper limits and the addition of one more Higgs
doublet,  which drives the flavor changing neutral currents
(FCNCs) and the LFV interactions at tree level, may cause to pull
the theoretical values of the BRs near to the experimental upper
limits. This is the case that the lepton FV is induced by the
internal new neutral Higgs bosons, $h^0$ and $A^0$, and the
strength of this violation is regulated by the Yukawa couplings,
appearing as free parameters which should be restricted by using
the experimental data. These decays were examined in the framework
of the SM with one more Higgs doublet \cite{Lee}-\cite{Atwood},
the so called two Higgs doublet model (2HDM), in
\cite{Iltan1,Iltan11, Diaz, IltanExtrDim, IltanLFVRS, Diaz2}.
Besides the theoretical calculations based on the 2HDM, they were
studied in the supersymmetric models
\cite{Barbieri1}-\cite{Barbieri7}, in a model independent way
\cite{Chang}, in the framework of 2HDM and the supersymmetric
model \cite{Paradisi} and, recently, in the SM including effective
operators coming from the possible unparticle effects
\cite{MuLinYan}-\cite{AndiHektor}.

Another possibility to enhance the numerical values of the BRs of
these processes is to consider the extra dimension which results
in the additional effects of the KK modes of the particles in the
loops, after the compactification. In the present work, we
consider the extended Higgs sector, the 2HDM, in the the RS1
background \cite{Rs1, Rs2}. The RS1 model is based on the curved
extra dimension and the corresponding metric reads
\begin{eqnarray}
ds^2=e^{-2\,\sigma}\,\eta_{\mu\nu}\,dx^\mu\,dx^\nu-dy^2\, ,
\label{metric1}
\end{eqnarray}
where $\sigma=k\,|y|$, $k$ is the bulk curvature constant, the
exponential $e^{-\sigma}$, with $y=R\,|\theta|$,  is the warp
factor. Here, the extra dimension is compactified onto $S^1/Z_2$
orbifold and $R$ is the compactification radius. The extra
dimension has two boundaries, the hidden (Planck) brane and the
visible (TeV) brane, with opposite and equal tensions. This choice
leads to the fact that the low energy effective theory has flat 4D
spacetime, even if the 5D cosmological constant is non vanishing.
In the RS1 background, the gravity is taken to be localized on the
hidden brane and to be extended into the bulk with varying
strength and the SM fields live in the visible brane. If some of
the SM fields are accessible to the extra dimension, the
phenomenology becomes richer and there are various work done in
the literature respecting such scenarios
\cite{Goldberger}-\cite{Moreau2}. If fermions are accessible to
the extra dimension and there is a Dirac mass term in the
Lagrangian, the fermion mass hierarchy can be explained. In this
case the fermion mass hierarchy comes from the possible fermion
field locations \cite{Pamoral2, Grossman, Huber, Huber2}. The
quark and lepton FV, which is based on the different locations of
the fermion fields in the extra dimension, is extensively studied
in \cite{KAgashe, EBlechman}. In these works, it is considered
that the FV is carried by the Yukawa interactions, coming from the
SM Higgs-fermion-fermion vertices. In \cite{Pree}, the high
precision measurements of top pair production at the ILC is
addressed  by considering that the fermions are localized in the
bulk of RS1 background. In recent works \cite{Moreau1, Moreau2},
the various experimental FCNC constraints and the electro weak
precision tests for the location parameters of the fermions in the
extra dimension are discussed. The other possibility is to
consider the localization of Higgs field in the extra dimension.
The brane localized mass terms for scalar fields are considered in
order to get small couplings of KK modes with the boundaries
\cite{Pamoral2} and  these mass terms result in that the zero mode
localized solution is obtained. In \cite{Huber3} the hierarchy of
fermion masses is analyzed by taking that the Higgs field has an
exponential profile around the TeV brane. \cite{Kogan} is devoted
to an extensive work on the bulk fields in various multi-brane
models.

In our work, we assume that the new Higgs doublet is accessible to
the extra dimension of RS1 background. First, we study the case
that the charged leptons are restricted to the 4D brane and,
second, we consider that the charged leptons are also localized in
the extra dimension. Notice that, in both cases, the gauge bosons
are necessarily accessible to the extra dimension.

The paper is organized as follows: In Section 2, we present the
BRs of LFV interactions $l_1\rightarrow l_2\gamma$ in the 2HDM, by
considering that the new Higgs doublet is localized in the extra
dimension of RS1 background. Section 3 is devoted to discussion
and our conclusions. In Appendix A, we study the construction of
new Higgs boson mass matrix. In Appendix B, we present the
amplitudes appearing in the calculation of the decay widths of the
radiative decays under consideration Appendix C is devoted to
calculation of the zero mode lepton fields and their KK modes.
\section{{LFV $l_1\rightarrow l_2\gamma$ decays in the Randall
Sundrum background with localized new Higgs boson}}
We start with the action for the new Higgs doublet $\phi$ (see for
example \cite{Goldberger}, \cite{Kogan} for a massive bulk scalar
field case),
\begin{eqnarray}
{\cal{S}}_\phi=\frac{1}{2}\,\int d^4x \int dy
\,\sqrt{g}\,\Big(g^{MN}\,(\partial_M
\phi)^\dagger\,\partial_N\,\phi +m^2_\phi\,\phi^\dagger\,\phi
 \Big) \label{actionphi} \, ,
\end{eqnarray}
where  $g=Det[g_{MN}]=e^{-8\,\sigma}$, $M,N=0,1,... ,4$. The
decomposition of the scalar doublet into KK modes
\begin{eqnarray}
\phi(x,y)=\sum_{n=0}^\infty\, \phi^{(n)}(x)\, f_n(y) \label{phiKK}
\end{eqnarray}
brings the action eq.(\ref{actionphi}) into form
\begin{eqnarray}
{\cal{S}}_S=\frac{1}{2}\,\sum_{n=0}^\infty\,\int d^4x
\,\Big(\eta^{\mu\nu}\,(\partial_\mu \phi_n
(x))^\dagger\,\partial_\nu\,\phi_n (x) +m_{n\,S}^2\,(\phi_n
(x))^\dagger\,\phi_n (x)
 \Big) \label{KKactionphi} \, ,
\end{eqnarray}
with the second order differential equation
\begin{eqnarray}
-e^{4\,\sigma}\,\frac{d}{dy}\,\Big(
e^{-4\,\sigma}\,\frac{d\,f_n(y)}{dy}\Big)+m_\phi^2\,f_n(y)=m_{n\,S}^2\,
e^{2\,\sigma}\,f_n(y) \label{phidiffeqn} \, ,
\end{eqnarray}
and the orthogonality relation
\begin{eqnarray}
\int_{-\pi\,R}^{\pi\,R} dy e^{-2\,\sigma}\,f^*_n(y)\,f_m(y)
=\delta_{nm}\label{phinorm} \, .
\end{eqnarray}
The choice of the mass term in eq.(\ref{actionphi})
\begin{eqnarray}
m^2_\phi=a\,(\frac{d\,\sigma}{dy})^2+b\,\frac{d^2\,\sigma}{dy^2}\,
, \label{mphi}
\end{eqnarray}
results in the differential equation in the bulk
\begin{eqnarray}
-e^{4\,\sigma}\,\frac{d}{dy}\,\Big(
e^{-4\,\sigma}\,\frac{d\,f_n(y)}{dy}\Big)+a\,k^2\,f_n(y)=m_{n\,S}^2\,
e^{2\,\sigma}\,f_n(y) \label{phidiffeqn2} \, ,
\end{eqnarray}
where $n=1,2,... .$. Now, the the boundary mass term\footnote{Here
the boundary mass terms have the same magnitude and the opposite
sign on the branes. The idea of brane localized mass terms has
been considered for scalar fields in \cite{Goldberger},
\cite{Pamoral2}.}
\begin{eqnarray}
m^2_{\phi, bound}=b\,\frac{d^2\,\sigma}{dy^2}\, ,\label{mphibound}
\end{eqnarray}
is considered in order to obtain zero mode Higgs doublet and this
term induces the boundary condition
\begin{eqnarray}
\Bigg( \frac{\partial \phi (x,y)}{\partial y}-b\,k\,\phi
(x,y)\Bigg)|_{y=0,\pi\,R}=0
 \,\, .\label{Boundary}
\end{eqnarray}
Notice that the non-vanishing zero mode can be obtained with the
fine tuning of the parameters\footnote{There is another
possibility of fine tuning of the parameters  $b$ and $a$ for the
non-vanishing zero mode, namely $b=2-\sqrt{4+a}$. However we
ignore this choice since it is not appropriate for the brane
localized fermion scenario and bulk fermion scenario with the
parameter set used (see discussion section for details).} $b$ and
$a$,
\begin{eqnarray}
b=2+\sqrt{4+a}
 \, ,\label{finetune}
\end{eqnarray}
and it reads
\begin{eqnarray}
f_0(y)=\frac{e^{b\,k\,y}}{\sqrt{\frac{e^{2\,(b-1)\,
k\,\pi\,R}-1}{(b-1)\,k}}}
 \,\, . \label{f0}
\end{eqnarray}
On the other hand, the KK mode Higgs doublet is obtained as
\begin{eqnarray}
f_n(y)=\frac{e^{2\,\sigma}}{N_{S\,n}}\, \Bigg( J_{\sqrt{4+a}}
(e^{\sigma}\,x_{nS})+\alpha_n\, Y_{\sqrt{4+a}}
(e^{\sigma}\,x_{nS})\Bigg)\label{nHmode} \, ,
\end{eqnarray}
where $N_{S\,n}$ is the normalization constant,
$x_{nS}=\frac{m_{n\,S}}{k}$
%
and $\alpha_n$ reads
\begin{eqnarray}
\alpha_n&=&
\frac{(2-b)\,J_{\sqrt{4+a}}(x_{nS})+x_{nS}\,J^\prime_{\sqrt{4+a}}
(x_{nS})}
{(b-2)\,Y_{\sqrt{4+a}}(x_{nS})-x_{nS}\,Y^\prime_{\sqrt{4+a}}(x_{nS})}
\label{alfn} \, .
\end{eqnarray}
Here, the functions $J_\beta(w)$ and $Y_\beta(w)$ are the Bessel
function of the first kind and of the second kind, respectively.
Finally, the mass spectrum of KK modes ($n=1,2,...$) are obtained
by using the boundary conditions at $y=0$ and $y=\pi\,R$ (see eq.(
\ref{Boundary})),
\begin{eqnarray}
m_{n\,S}\simeq
(n+\frac{1}{2}\,(b-2)-\frac{3}{4})\,\pi\,k\,e^{-k\,\pi\,R} \,\,
,\label{mn}
\end{eqnarray}
for $k\,e^{-k\,\pi\,R}\ll m_{n\,S}\ll k$.

At this stage, we consider two  possibilities for the charged
leptons :
\begin{itemize}
\item they are restricted to the 4D brane
\item they are localized in the extra dimension.
\end{itemize}
\subsection{The charged leptons restricted to the brane} \label{sub1}
The LFV interactions are driven by the part of the action
\begin{eqnarray}
{\cal{S}}_{Y}= \int d^5x \sqrt{g} \,\Bigg( \xi^{E}_{5\,
ij}\,\bar{l}_{i L} \phi_{2} E_{j R} + h.c. \Bigg)\, \delta(y-\pi
R) \,\,\, , \label{yukawalagrangian}
\end{eqnarray}
where $L$ and $R$ denote chiral projections $L(R)=1/2(1\mp
\gamma_5)$, $\phi_{2}$ is the new scalar doublet, $l_{i L}$ ($E_{j
R}$) are lepton doublets (singlets), $\xi^{E}_{5\,ij}$, with
family indices $i,j$ , are the Yukawa couplings in five
dimensions, which are responsible for the flavor violating
interactions in the leptonic sector. Here, we assume that the
Higgs doublet $\phi_1$ lives on the visible brane and it has
non-zero vacuum expectation value in order to ensure the ordinary
masses of the gauge fields and the fermions. On the other hand the
second doublet, which is accessible to the extra dimension, has no
vacuum expectation value\footnote{ Here we take the Higgs
potential as
\begin{eqnarray}
V(\phi_1, \phi_2 )=c_1 (\phi_1^+ \phi_1-v^2/2)^2+ c_2 (\phi_2^+
\phi_2)^2 + c_3 [(\phi_1^+ \phi_1) (\phi_2^+ \phi_2)-(\phi_1^+
\phi_2)(\phi_2^+ \phi_1)]+ c_4 [Re(\phi_1^+ \phi_2)]^2 + c_5
[Im(\phi_1^+ \phi_2)]^2 \, . \label{potential} \nonumber
\end{eqnarray}
This choice leads to no tree level mixing between the CP even
neutral Higgs bosons, namely $H^0$ and $H^1$. Therefore, the SM
particles (new particles) are collected in the first (second)
doublet and $H^1$, $H^2$ are obtained as the mass eigenstates
$h^0$ and $A^0$ respectively. Notice that, in general, the mixing
between the CP even neutral Higgs bosons can exist in the loop
level when one considers the quantum corrections.}:
\begin{eqnarray}
\phi_{1}=\frac{1}{\sqrt{2}}\left[\left(\begin{array}{c c}
0\\v+H^{0}\end{array}\right)\; + \left(\begin{array}{c c} \sqrt{2}
\chi^{+}\\ i \chi^{0}\end{array}\right) \right]\, ;
\phi_{2}=\frac{1}{\sqrt{2}}\left(\begin{array}{c c} \sqrt{2}
H^{+}\\ H^1+i H^2 \end{array}\right) \,\, , \label{choice}
\end{eqnarray}
and
\begin{eqnarray}
<\phi_{1}>=\frac{1}{\sqrt{2}}\left(\begin{array}{c c}
0\\v\end{array}\right) \,  \, ; <\phi_{2}>=0 \,\, .
\label{choice2}
\end{eqnarray}
The new Higgs doublet $\phi_{2}$ is expanded into KK modes after
the compactification of the extra dimension as given in
eq.(\ref{phiKK}) and the zero (KK) mode Higgs fields are obtained
by imposing the fine tuning condition in eq.(\ref{finetune}). On
the other hand, after the electro weak breaking, the SM Higgs
acquires the vacuum expectation value eq.(\ref{choice2}) and there
appears mixing between zero mode and KK mode new Higgs bosons.
However, we do not take into account the additional effects coming
from this mixing since they are suppressed (see Appendix A for
detail).

For the effective Yukawa coupling $\xi^{E}_{ij}$ we integrate out
the Yukawa interaction eq.(\ref{yukawalagrangian}) over the fifth
dimension by taking the zero mode neutral Higgs fields $S=h^0,A^0$
:
\begin{eqnarray}
\xi^{E}_{ij}=V^0_{ij}\, \xi^{E}_{5\, ij}\label{yukawa0} \, ,
\end{eqnarray}
where
\begin{eqnarray}
V^0_{ij}&=&\int_{-\pi\,R}^{\pi\,R}\,dy\, e^{-4\,\sigma}\,f_0(y)
\,\delta(y-\pi R) \nonumber \\ &=&
\frac{e^{(b-4)\,k\,\pi\,R}}{\sqrt{\frac{e^{2\,(b-1)\,k\,\pi\,R}-1}
{k\,(b-1)}}} \label{V0ij} \, .
\end{eqnarray}
Here it is assumed that the coupling $\xi^{E}_{5\, ij}$ is flavor
dependent and it is regulated in such a way that the overall
quantity $V^0_{ij}\,\xi^{E}_{5\, ij}$ is pointed to the chosen
numerical value of $\xi^{E}_{ij}$.

The effective Yukawa coupling inducing the tree level interaction
among the KK mode Higgs and charged leptons reads
\begin{eqnarray}
\xi^{E\,n}_{ij}= V^{S\,n}_{ij}\,\xi^{E}_{5\, ij}
\label{ksinijilk} \, ,
\end{eqnarray}
where
\begin{eqnarray}
V^{S\,n}_{ij}= \frac{e^{-2\,k\,\pi\,R}}{N_{S\,n}}
\,\Big(J_{b-2}(e^{k\,\pi\,R}\,x_{nS})+\alpha_n\, Y_{b-2}(
e^{k\,\pi\,R}\,x_{nS})\Big) \label{VSn} \, .
\end{eqnarray}
Here $N_{S\,n}$ is the normalization constant (see
eq.(\ref{nHmode})) and $b$ ($\alpha_n$) is defined in
eqs.(\ref{finetune}) ((\ref{alfn})). Finally, the effective Yukawa
coupling $\xi^{E\,n}_{ij}$ is obtained as
\begin{eqnarray}
\xi^{E\,n}_{ij}=\frac{V^n_{S\,ij}}{V^0_{ij}}\,\xi^{E}_{ij}=
\frac{e^{(2-b)\,k\,\pi\,R}\,\sqrt{\frac{e^{2\,(b-1)\,k\,\pi\,R}-1)}
{k\,(b-1)}}}{N_{S\,n}}
\,\Big(J_{b-2}(e^{k\,\pi\,R}\,x_{nS})+\alpha_n\, Y_{b-2}(
e^{k\,\pi\,R}\,x_{nS})\Big)\,\xi^{E}_{ij} \label{ksinij} \, .
\end{eqnarray}

\begin{figure}[htb]
\vskip 6.0truein \centering \epsfxsize=3.8in
\leavevmode\epsffile{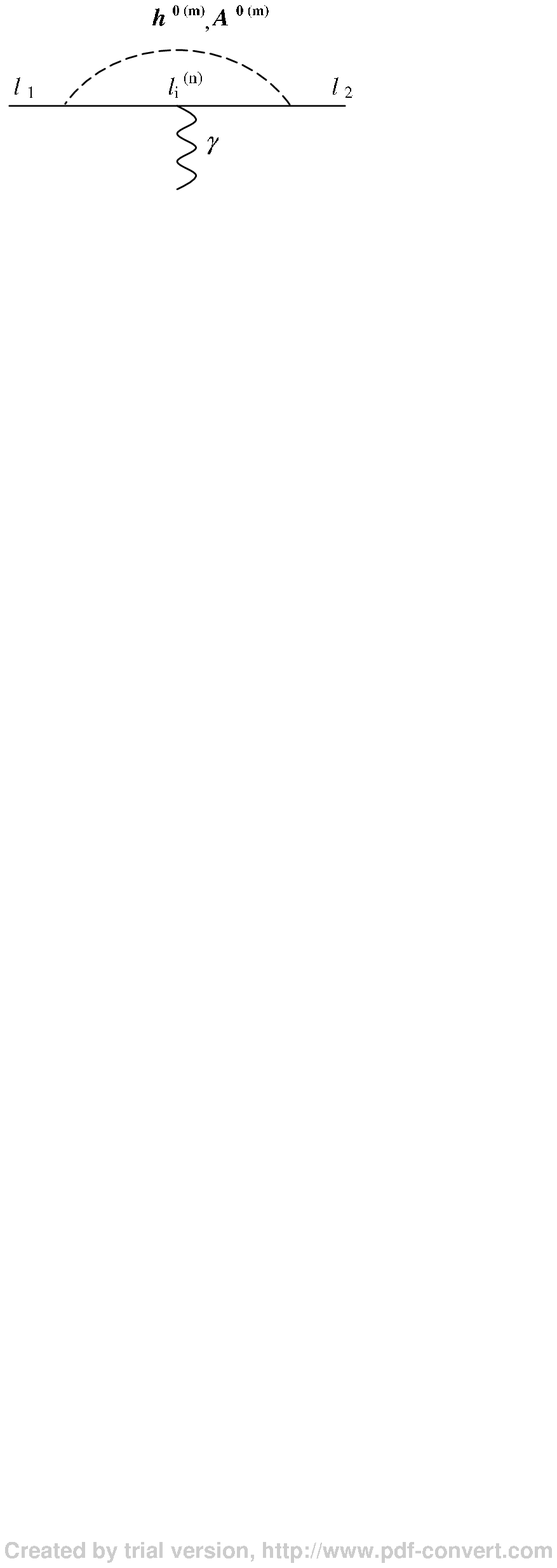} \vskip -9.0truein \caption[]{One
loop diagrams contribute to $l_1\rightarrow l_2 \gamma$ decay  due
to the zero mode (KK mode) leptons and Higgs fields in the 2HDM.}
\label{fig1}
\end{figure}

Now, we present the decay widths of the LFV $l_1\rightarrow
l_2\gamma$ decays, including the KK modes of new neutral Higgs
fields. Since these decays exist at least in the one loop level,
there appear the logarithmic divergences in the calculations. In
order to eliminate these divergences, we follow  the on-shell
renormalization scheme. In this scheme, the self energy diagrams
can be written in the form $\sum(p)=(\hat{p}-m_{l_1})\bar{\sum}(p)
(\hat{p}-m_{l_2})\, , $ which results in that these diagrams do
not contribute for on-shell leptons and, only, the vertex diagrams
(see Fig.\ref{fig1}) contribute\footnote{This is the case that the
divergences can be eliminated by introducing a counter term
$V^{C}_{\mu}$ with the relation
$V^{Ren}_{\mu}=V^{0}_{\mu}+V^{C}_{\mu} \, , $ where
$V^{Ren}_{\mu}$ ($V^{0}_{\mu}$) is the renormalized (bare) vertex
and by using the gauge invariance $k^{\mu} V^{Ren}_{\mu}=0$. Here,
$k^\mu$ is the four momentum vector of the outgoing photon.}.
Taking only tau lepton for the internal line\footnote{We take into
account only the internal tau lepton contribution since we respect
the idea that the couplings $\bar{\xi}^{E}_{N, ij}$ ($i,j=e,\mu$),
are small compared to $\bar{\xi}^{E}_{N,\tau\, i}$
$(i=e,\mu,\tau)$, due to the possible proportionality of them to
the masses of leptons under consideration in the vertices. Here,
we use the dimensionful coupling $\bar{\xi}^{E}_{N,ij}$ with the
definition $\xi^{E}_{N,ij}=\sqrt{\frac{4\, G_F}{\sqrt{2}}}\,
\bar{\xi}^{E}_{N,ij}$ where N denotes the word "neutral".}, the
decay width $\Gamma$ reads
\begin{eqnarray}
\Gamma (l_1\rightarrow l_2\gamma)=c_1(|A_1|^2+|A_2|^2)\,\, ,
\label{DWmuegam}
\end{eqnarray}
where
\begin{eqnarray}
A_1&=&A^0_1+A^{S\, KK}_1 \nonumber \, , \\
A_2&=&A^0_2+A^{S\, KK}_2\, . \label{A1A1S}
\end{eqnarray}
For the explicit expression of these amplitudes see Appendix B.
\subsection{The charged leptons localized in the extra dimension}
\label{sub2}
The part of the action which drives the LFV interactions in this
case reads
\begin{eqnarray}
{\cal{S}}_{Y}= \int d^5x \sqrt{g} \,\Bigg( \xi^{E}_{5\,
ij}\,\bar{l}_{i L} \phi_{2} E_{j R} + h.c. \Bigg)\,  ,
\label{yukawalagrangian2}
\end{eqnarray}
where  $l_{i L}$ ($E_{j R}$) are lepton doublets (singlets) which
are localized in the extra dimension. The addition of Dirac mass
term to the lagrangian of bulk fermions causes this localization
\cite{Hisano, Hewett, Pamoral2, Huber4, Grossman, Huber2,Huber3}.
Since the combination $\bar{\psi}\psi$ is odd due to the two
possible transformation properties of fermions under the orbifold
$Z_2$ symmetry, $Z_2\psi=\pm\gamma_5 \psi$, in order to construct
the $Z_2$ invariant mass term, one needs  $Z_2$ odd scalar field
to be coupled. This discussion leads to the mass term
\begin{eqnarray}
{\cal{S}}_m=-\int d^4x \int dy \,\sqrt{g}\,m(y)\,\bar{\psi}\psi
\label{massterm} \, ,
\end{eqnarray}
where $m(y)=m\frac{\sigma'(y)}{k}$ with
$\sigma'(y)=\frac{d\sigma}{dy}$. With the help of the given mass
term the localized zero mode leptons are obtained. We present the
construction of the zero mode and KK mode leptons in the Appendix
C extensively.

For the effective Yukawa coupling $\xi^{E}_{ij}$, similar to the
previous case, we integrate out the Yukawa interaction
eq.(\ref{yukawalagrangian2}) over the fifth dimension. By taking
the zero mode lepton doublets, singlets (see eq. (\ref{0mode}))
and neutral Higgs fields $S=h^0,A^0$ (see eq.(\ref{f0})), we get
\begin{eqnarray} \xi^{E}_{ij}\,((\xi^{E}_{ij})^\dagger)=
V^{00}_{S\,RL(LR)\,ij}\, \xi^{E}_{5\,
ij}\,((\xi^{E}_{5\,ij})^\dagger) \label{yukawa0RL} \, ,
\end{eqnarray}
where
\begin{eqnarray}
V^{00}_{S\,RL\,ij}&=&\frac{1}{2\,\pi\,R}\,\int_{-\pi\,R}^{\pi\,R}\,dy\,
\chi_{iR0}(y)\,\chi_{jL0}(y)\,f_0(y) \nonumber \\ &=& \frac{ \Big(
1-e^{(b-r_{iR}-r_{jL})\,k\,\pi\,R}\Big)}
{(r_{iR}+r_{jL}-b)\,\sqrt{\frac{
1-e^{(1-2\,r_{iR})\,k\,\pi\,R}}{(2\,r_{iR}-1)}}\,\sqrt{\frac{
1-e^{(1-2\,r_{jL})\,k\,\pi\,R}}{(2\,r_{jL}-1)}}\,
\sqrt{\frac{1-e^{2\,(b-1)\,k\,\pi\,R}}{k\,(1-b)}}}
\label{yukawa0LR} \, .
\end{eqnarray}
Here, similar to the previous scenario, the coupling $\xi^{E}_{5\,
ij}$ in five dimension is flavor dependent and it is regulated in
such a way that the overall quantity
$V^0_{RL\,(LR)\,ij}\,\xi^{E}_{5\, ij}$ is pointed to the chosen
numerical value of
$\xi^{E}_{ij}\,\Big((\xi^{E}_{ij})^\dagger\Big)$. This is the case
that the hierarchy of new Yukawa couplings, describing the tree
level Higgs zero mode($S^{(0)}$)-lepton zero mode
($l^{(0)}$)-lepton zero mode ($l^{(0)}$) interaction, is not
related to the Higgs field and lepton field locations.

The effective Yukawa coupling which drives the tree level KK mode
Higgs ($S^{(n)}$)-$l^{(0)}$-$l^{(0)}$ interaction is
\begin{eqnarray}
\xi^{E\,n\,0}_{ij}\,((\xi^{E\,n\,0}_{ij})^\dagger)=
V^{n0}_{S\,RL(LR)\,ij}\,\xi^{E}_{5\,
ij}\,((\xi^{E}_{5\,ij})^\dagger) \label{ksinLR0} \, ,
\end{eqnarray}
where
\begin{eqnarray}
V^{n0}_{S\,RL\,ij}&=&
\frac{\int_{-\pi\,R}^{\pi\,R}\,dy\,e^{(2-r_{iR}-
r_{jL})\,\sigma}\,\Big(J_{b-2}(e^{\sigma}\,x_{nS})+\alpha_n\,
Y_{b-2}( e^{\sigma}\,x_{nS}) \Big)}{N_{S\,n}\,\sqrt{\frac{
1-e^{(1-2\,r_{iR})\,k\,\pi\,R}}{k\,(2\,r_{iR}-1)}}\,\sqrt{\frac{
1-e^{(1-2\,r_{jL})\,k\,\pi\,R}}{k\,(2\,r_{jL}-1)}}}
\label{yukawanLR} \, .
\end{eqnarray}
Using the eqs. (\ref{ksinLR0}) and (\ref{yukawanLR}), the
effective Yukawa coupling $\xi^{E\,n\,0}_{ij}$ is obtained as
\begin{eqnarray}
\xi^{E\,n\,0}_{ij}&=&\frac{V^{n0}_{S\,RL\,ij}}{V^{00}_{S\,RL\,ij}}\,
\xi^{E}_{ij}=
\frac{(r_{iR}+r_{jL}-b)\,\sqrt{\frac{k\,(e^{2\,(b-1)\,k\,\pi\,R}-1)}
{(b-1)}}}{N_{S\,n}\,\Big(1-e^{(b-r_{iR}-r_{jL})\,k\,\pi\,R}\Big)}
\nonumber \\ &\times& \int_{-\pi\,R}^{\pi\,R}\,dy\,e^{(2-r_{iR}-
r_{jL})\,\sigma}\,\Big(J_{b-2}(e^{\sigma}\,x_{nS})+\alpha_n\,
Y_{b-2}( e^{\sigma}\,x_{nS})\Big)\,\xi^{E}_{ij} \label{ksin0ij} \,
.
\end{eqnarray}
$S^{(0)}-l^{(0)}-l^{(n)}$ vertex drives another possible tree
level interaction appearing in the loop calculations and the
corresponding is effective Yukawa coupling reads
\begin{eqnarray}
\xi^{E\,0\,n}_{ij}\,\Big((\xi^{E\,0\,n}_{ij})^\dagger \Big)=
V^{0n}_{S\,RL\,(LR)\,ij}\,\xi^{E}_{5\, ij}\,\Big((\xi^{E}_{5\,
ij})^\dagger\Big) \label{ksi0LRn} \, ,
\end{eqnarray}
where
\begin{eqnarray}
V^{0n}_{S\,RL\,(LR)\,ij}=\frac{N_{Ln\,(Rn)}\,\int_{-\pi\,R}^{\pi\,R}\,dy\,
e^{(b-r_{iR\,(iL)}+\frac{1}{2})\, \sigma}\,\Bigg(
J_{\frac{1}{2}\mp r_{jL\,(jR)}} (e^\sigma\,x_{nL(R)})+
c_{L\,(R)}\, Y_{\frac{1}{2}\mp
r_{jL\,(jR)}}(e^\sigma\,x_{nL(R)})\Bigg)}{\pi\,R\,\sqrt{\frac{
1-e^{(1-2\,r_{iR\,(iL)})\,k\,\pi\,R}}{k\,\pi\,R\,(2\,r_{iR\,(iL)}-1)}}
\,\sqrt{\frac{e^{2\,(b-1)\,k\,\pi\,R}-1}{(b-1)\,k}}}\, .\nonumber \\
\label{yukawaonLR}
\end{eqnarray}
Here the parameters $x_{nR(L)}$, $c_R(L)$, the lepton localization
parameters $r_{iR\,(iL)}$ and the normalization constant
$N_{R(L)n}$ are given in Appendix C. By using the eqs.
(\ref{ksi0LRn}) and (\ref{yukawaonLR}) we get the effective Yukawa
coupling $\xi^{E\,0\,n}_{ij}\,\Big((\xi^{E\,0\,n}_{ij})^\dagger
\Big)$  as
\begin{eqnarray}
\xi^{E\,0\,n}_{ij}\!\!\!\! \!\!\!\!\!\!\!\!\!\!\!& &
\Big((\xi^{E\,0\,n}_{ij})^\dagger \Big)=
\frac{V^{0n}_{S\,RL\,(LR)\,ij}}{V^{00}_{S\,RL(LR)\,ij}}\,
\xi^{E}_{ij}\,\Big( (\xi^{E}_{ij})^\dagger \Big)=N_{Ln\,(Rn)}\,
\,\sqrt{\frac{
k\,(1-e^{(1-2\,r_{jL\,(jR)})\,k\,\pi\,R})}{\pi\,R\,(2\,r_{jL\,(iR)}-1)}}
(r_{iR\,(iL)}+r_{jL\,(jR)}-b)
\nonumber \\
&\times& \frac{\int_{-\pi\,R}^{\pi\,R}\,dy\,
e^{(b-r_{iR\,(iL)}+\frac{1}{2})\, \sigma}\,\Bigg(
J_{\frac{1}{2}\mp r_{jL\,(jR)}} (e^\sigma\,x_{nL(R)})+
c_{L\,(R)}\, Y_{\frac{1}{2}\mp
r_{jL\,(jR)}}(e^\sigma\,x_{nL(R)})\Bigg)}
{\Big(1-e^{(b-r_{iR\,(iL)}-r_{jL\,(jR)})\,k\,\pi\,R}\Big)}
\,\xi^{E}_{ij}\,\Big((\xi^{E}_{ij})^\dagger\Big)\, . \nonumber \\
\label{ksin0nij}
\end{eqnarray}
Finally, the tree level $S^{(m)}-l^{(0)}-l^{(n)}$ interaction is
carried by the effective Yukawa coupling $\xi^{E\,m\,n}_{ij}$ and
it reads
\begin{eqnarray}
\xi^{E\,m\,n}_{ij}\,\Big((\xi^{E\,m\,n}_{ij})^\dagger \Big)=
V^{mn}_{S\,RL\,(LR)\,ij}\,\xi^{E}_{5\, ij}\,\Big((\xi^{E}_{5\,
ij})^\dagger\Big) \label{ksimLRn} \, ,
\end{eqnarray}
with
\begin{eqnarray}
V^{mn}_{S\,RL\,(LR)\,ij}=\frac{N_{Ln\,(Rn)}}{N_{S\,m}\,\pi\,R\,
\sqrt{\frac{1-e^{(1-2\,r_{iR\,(iL)})\,k\,\pi\,R}}{k\,\pi\,R\,(2\,r_{iR\,(iL)}
-1)}}}\!\!\!\! \!\!\!\!\!\!\!\!& &\int_{-\pi\,R}^{\pi\,R}\,dy\,
e^{(\frac{5}{2}-r_{iR\,(iL)})\, \sigma}\, \Bigg(J_{b-2}
(e^\sigma\,x_{mS})+\alpha_n
\,Y_{b-2}(e^\sigma\,x_{mS})\Bigg)\nonumber \\ &\times& \Bigg(
J_{\frac{1}{2}\mp r_{jL\,(jR)}} (e^\sigma\,x_{nL(R)})+
c_{L\,(R)}\, Y_{\frac{1}{2}\mp
r_{jL\,(jR)}}(e^\sigma\,x_{nL(R)})\Bigg)\, , \nonumber
\\ \label{yukawamnLR}
\end{eqnarray}
and, therefore, we get
\begin{eqnarray}
\xi^{E\,m\,n}_{ij}\!\!\!\! \!\!\!\!\!\!\!\!\!\!\!& &
\Big((\xi^{E\,m\,n}_{ij})^\dagger \Big)=
\frac{V^{mn}_{S\,RL\,(LR)\,ij}}{V^{00}_{S\,RL(LR)\,ij}}\,
\xi^{E}_{ij}\,\Big( (\xi^{E}_{ij})^\dagger \Big)\nonumber \\ &=&
\frac{N_{Ln\,(Rn)}\,
\sqrt{\frac{1-e^{(1-2\,r_{jL\,(jR)})\,k\,\pi\,R}}{\pi\,R\,(2\,r_{jL\,(jR)}
-1)}}\, \sqrt{\frac{e^{2\,(b-1)\,k\,\pi\,R}-1}{(b-1)}}
(r_{iR\,(iL)}+r_{jL\,(jR)}-b)}
{N_{S\,m}\,\Big(1-e^{(b-r_{iR\,(iL)}-r_{jL\,(jR)})\,k\,\pi\,R}\Big)}
\nonumber \\&\times& \int_{-\pi\,R}^{\pi\,R}\,dy\,
e^{(\frac{5}{2}-r_{iR\,(iL)})\, \sigma}\, \Bigg(J_{b-2}
(e^\sigma\,x_{mS})+\alpha_n
\,Y_{b-2}(e^\sigma\,x_{mS})\Bigg)\nonumber \\ &\times& \Bigg(
J_{\frac{1}{2}\mp r_{jL\,(jR)}} (e^\sigma\,x_{nL(R)})+
c_{L\,(R)}\, Y_{\frac{1}{2}\mp
r_{jL\,(jR)}}(e^\sigma\,x_{nL(R)})\Bigg)
\,\xi^{E}_{ij}\,\Big((\xi^{E}_{ij})^\dagger\Big) \, .\nonumber \\
\label{ksimn}
\end{eqnarray}

The decay widths of the LFV $\mu\rightarrow e\gamma$,
$\tau\rightarrow e\gamma$ and $\tau\rightarrow \mu\gamma$ decays
are calculated by using on-shell renormalization scheme (see the
section \ref{sub1}) and we have
\begin{eqnarray}
\Gamma (l_1\rightarrow l_2\gamma)=c_1(|A_1|^2+|A_2|^2)\,,
\label{DWmuegam2}
\end{eqnarray}
with
\begin{eqnarray}
A_1&=&A^0_1+A^{S\, KK}_1+A^{l\, KK}_1+A^{S,\, l\, KK}_1
\, , \nonumber \\
A_2&=&A^0_2+A^{S\, KK}_2+A^{l\, KK}_2+A^{S,\, l\, KK}_2 \, .
\label{A120SlKK}
\end{eqnarray}
Notice that we present the explicit expression of the amplitudes
$A^0_{1\,(2)}$, $A^{S\, KK}_{1\,(2)}$ and $A^{S,\, l\,
KK}_{1\,(2)}$ in Appendix B.
\newpage
\section{Discussion}
The Yukawa interactions coming from lepton-lepton-$S$ vertices
drive the radiative LFV $l_1\rightarrow l_2\gamma$
decays\footnote{Here, we do not take into account the internal
neutrino mediation due to their weak contribution to the $BRs$ of
the processes we study and, therefore, we assume that
the lepton FV comes from the internal new neutral Higgs bosons,
$h^0$ and $A^0$. Notice that we ignored the possible restrictions
coming from the hadronic decays.}, and their strengths are
regulated by the Yukawa couplings which are free parameters of the
model used. In the present work, we study these LFV decays in the
RS1 background and we assume that the new Higgs doublet and the
gauge fields are accessible to the extra dimension. Here, in order
to obtain zero mode Higgs doublet, one considers the boundary mass
term (see eq.(\ref{mphibound})) and impose the fine tuning
$b=2+\sqrt{4+a}$ (eq.(\ref{finetune})) of the parameters $b$ and
$a$ which regulates to the boundary and bulk mass terms. Finally,
the zero mode new Higgs doublet is obtained as an exponential
function of the parameter $b$ (eq.(\ref{f0})) and it is highly
localized around the visible brane.
The choice $b=2-\sqrt{4+a}$ is also possible for the non-vanishing
zero mode. However, we do not take this possibility into account
because of the following reason: If the fermions are localized on
the 4D brane, the overall quantity $V^0_{ij}\,\xi^{E}_{5\, ij}$ is
fixed to the chosen numerical value of $\xi^{E}_{ij}$ with the
assumption that the coupling $\xi^{E}_{5\, ij}$ is flavor
dependent and appropriately regulated.  For $b=2-\sqrt{4+a}$ the
coupling $V^0_{ij}$ is a number of orders smaller compared to
$b=2+\sqrt{4+a}$ and, since this term appear in the denominator of
the coupling $\xi^{E\,n}$ according to our definition,
$\xi^{E\,n}$ exceeds the range of perturbative calculation. Notice
that the coupling $V^{S\,n}_{ij}$ is not so much sensitive to the
parameter $b$. In the case of bulk fermions, the choice
$b=2-\sqrt{4+a}$ results in extremely large coupling which breaks
the perturbative upper limit (negligible coupling which causes
weak sensitivity to KK mode contributions) for set II (set I) . 

As a first attempt we assume that the leptons are restricted to
the 4D brane. In this case the contribution of the extra dimension
is due to the new Higgs KK modes which appear in the internal line
of the loop with the modified Yukawa couplings
(eq.(\ref{ksinij})). Second, we consider that the leptons are also
localized in the extra dimension. We follow the idea that the
localization of the lepton fields in the extra dimension occurs
with the addition of a Dirac mass term $m_l=r \sigma'$ with
$\sigma=k\,|y|$ (eq.(\ref{massterm})). In this case, the right and
left handed lepton zero modes (eq.(\ref{0mode})) are chosen to
locate at different positions in the extra dimension in order to
explain different flavor mass hierarchy. In the scenario we choose
the contribution of the extra dimension is coming from the new
Higgs KK modes and the lepton KK modes appearing in the internal
line of the loop. The FV is carried by the new Yukawa couplings
which are fixed to an appropriate number, respecting the current
measurements and the location parameters of leptons are
responsible for the lepton mass hierarchy. This choice makes the
constraints coming from various LFV processes to be more relaxed.
Here, we consider two different set of locations of charged
leptons in order to obtain the masses of different
flavors\footnote{The gauge sector is necessarily lives in the
extra dimension and their KK modes appear after the
compactification of the extra dimension. The different fermion
locations can induce additional FCNC effects at tree level due to
the couplings of neutral gauge KK modes-leptons and they should be
suppressed even for low KK masses, by choosing the location
parameters $r_L$ ($r_R$) appropriately. In the set of location
parameters we use (Table \ref{set}), we verify the various
experimental FCNC constraints with KK neutral gauge boson masses
as low as few TeVs (see the similar the set of location parameters
and the discussion given in \cite{Moreau1, Moreau2}.)}. In the
first set (Table \ref{set}), we consider the left and right handed
fields having the same location in the extra dimension. In the
second, we choose the left handed charged lepton locations as the
same for each flavor, and we estimate the right handed ones by
respecting the current charged lepton masses. For the second set,
we observe that the BRs of the decays under consideration enhance
since the KK mode couplings to the new Higgs scalars, which are
highly localized near the visible brane, become stronger if the
left handed lepton field is near to this brane. For the effective
Yukawa couplings in four dimension we choose that
$\bar{\xi}^{E}_{N,ij},\, i,j=e,\mu $ are smaller compared to
$\bar{\xi}^{E}_{N,\tau\, i}\, i=e,\mu,\tau$, since latter ones
contain heavy flavor and we assume that, in four dimensions, the
couplings $\bar{\xi}^{E}_{N,ij}$ is symmetric with respect to the
indices $i$ and $j$. Furthermore, the curvature parameter $k$ and
the compactification radius $R$ are among the free parameters of
the theory. Here, we take $k\,R=10.83$ and consider in the region
$10^{17}\,(GeV)\leq k \leq 10^{18} \,(GeV)$ (see the discussion in
Appendix C and \cite{Huber2}). Throughout our calculations we use
the input values given in Table (\ref{input}).
\begin{table}[h]
        \begin{center}
        \begin{tabular}{|l|l|}
        \hline
        \multicolumn{1}{|c|}{Parameter} &
                \multicolumn{1}{|c|}{Value}     \\
        \hline \hline
        $m_{\mu}$                   & $0.106$ (GeV) \\
        $m_{\tau}$                  & $1.78$ (GeV) \\
        $m_{h^0}$           & $100$   (GeV)  \\
        $m_{A^0}$           & $200$   (GeV)  \\
        $G_F$             & $1.16637 10^{-5} (GeV^{-2})$  \\
        \hline
        \end{tabular}
        \end{center}
\caption{The values of the input parameters used in the numerical
          calculations.}
\label{input}
\end{table}
%

In the case that the leptons live on the 4D brane
the contribution of the Higgs boson KK modes is negligible for the
decays under consideration
The weakness of the new contribution is due to the tiny ratio
$z_{Sn}$ appearing in the expression (eq.(\ref{A1A2SKK})) which
represents the additional effects to the amplitudes.

Now we analyze the case that the leptons are also accessible to
the extra dimension.

\begin{figure}[htb]
\vskip -3.0truein \centering \epsfxsize=6.8in
\leavevmode\epsffile{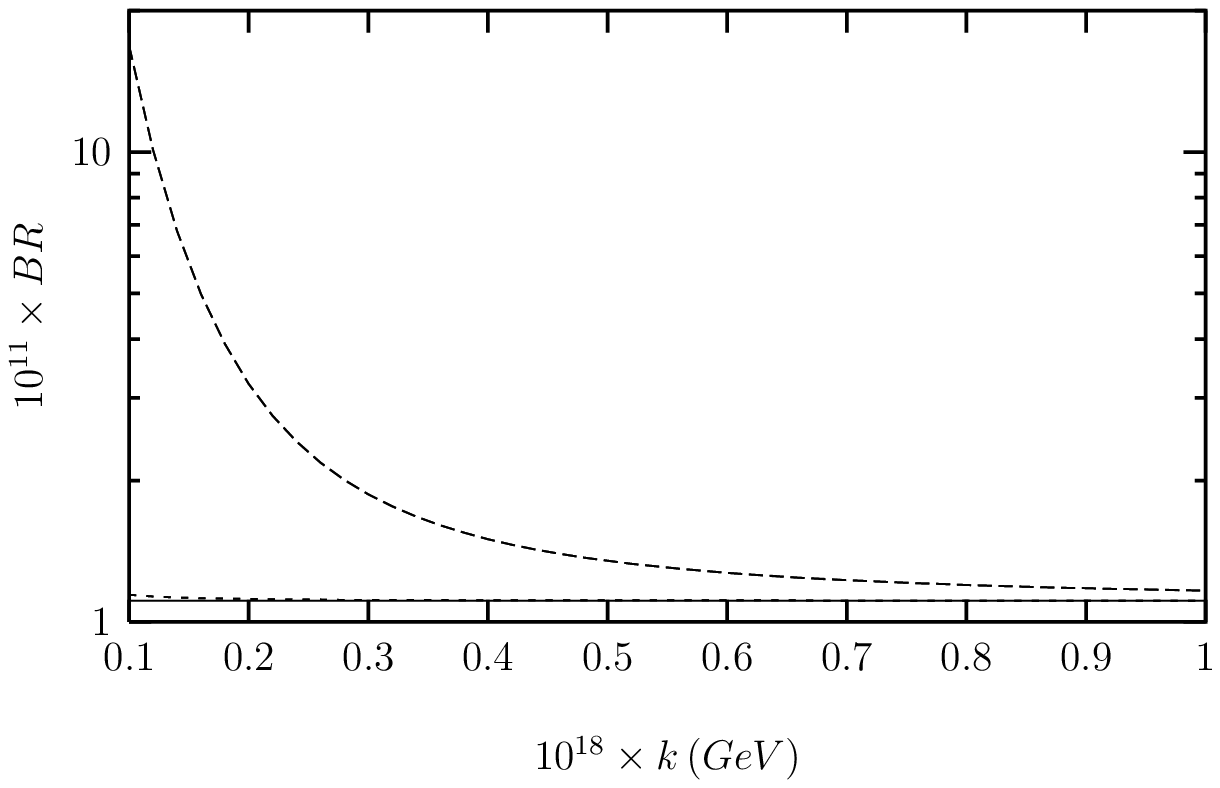} \vskip -3.0truein
\caption[]{$k$ dependence of the BR($\mu\rightarrow e \gamma$) for
$\bar{\xi}^{E}_{N,\tau e}=0.01\, GeV$,
$\bar{\xi}^{E}_{N,\tau\mu}=1.0\,GeV$. Here the solid (dashed,
short dashed) line represents the BR without KK modes of leptons
and new Higgs bosons (with KK modes of leptons and new Higgs
bosons for lepton location set II, set I), for $a=0.01$ and
$0.1$.} \label{BRleptBulkmuegamk}
\end{figure}

\begin{figure}[htb]
\vskip -3.0truein \centering \epsfxsize=6.8in
\leavevmode\epsffile{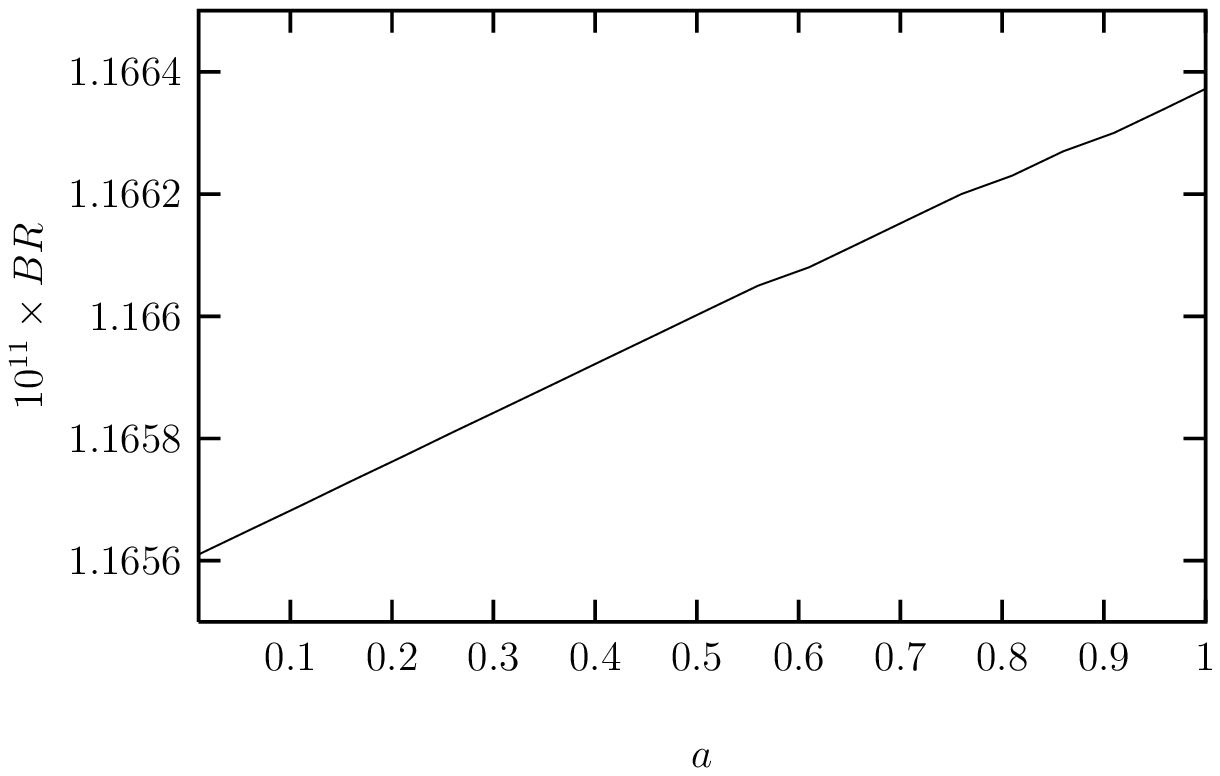} \vskip -3.0truein
\caption[]{$a$ dependence of the BR($\mu\rightarrow e \gamma$) for
$\bar{\xi}^{E}_{N,\tau e}=0.01\, GeV$,
$\bar{\xi}^{E}_{N,\tau\mu}=1.0\,GeV$, for the lepton location set
II and $k = 10^{18}\,GeV$.} \label{BRleptBulkmuegamabehv}
\end{figure}

Fig.\ref{BRleptBulkmuegamk} represents the parameter $k$
dependence of the BR of the  LFV $\mu\rightarrow e \gamma$ decay
for $\bar{\xi}^{E}_{N,\tau e}=0.01\, GeV$,
$\bar{\xi}^{E}_{N,\tau\mu}=1.0\,GeV$. Here the solid (dashed,
short dashed) line represents the BR without KK modes of leptons
and new Higgs bosons (with KK modes of leptons and new Higgs
bosons for lepton location set II, set I), for $a=0.01$ and
$0.1$\footnote{For $a=0.01$ and $a=0.1$ the curves almost
coincide}. It is observed that the BR ($\mu\rightarrow e \gamma$)
is of the order of $10^{-11}$ without the internal lepton and new
Higgs boson KK mode contributions. The addition of these KK modes
result in that the BR enhances almost $2\times$ one order, for the
lepton location set II, especially for the small values of the
parameter $k$. For the set I, the enhancement of the BR is
negligible. On the other hand, the BRs are weakly sensitive to the
parameter $a$ which plays a crucial role in the localization of
new Higgs bosons. We present the parameter $a$ dependence of the
BR ($\mu\rightarrow e \gamma$) for the lepton location set II, in
Fig.\ref{BRleptBulkmuegamabehv} for $k = 10^{18}\,GeV$. This
figure shows that the enhancement of the BR ($\mu\rightarrow e
\gamma$) is of the order of $\sim 0.1\%$ in the range of $a$,
$0.01 \leq a \leq 1.0$. This is a negligible enhancement which can
not be determined. On the other hand the enhancement in the case
of set II is due to the fact that the left handed leptons (KK
modes) are near to the visible brane and their couplings to the
new Higgs bosons, which are localized near the visible brane,
become stronger.

\begin{figure}[htb]
\vskip -3.0truein \centering \epsfxsize=6.8in
\leavevmode\epsffile{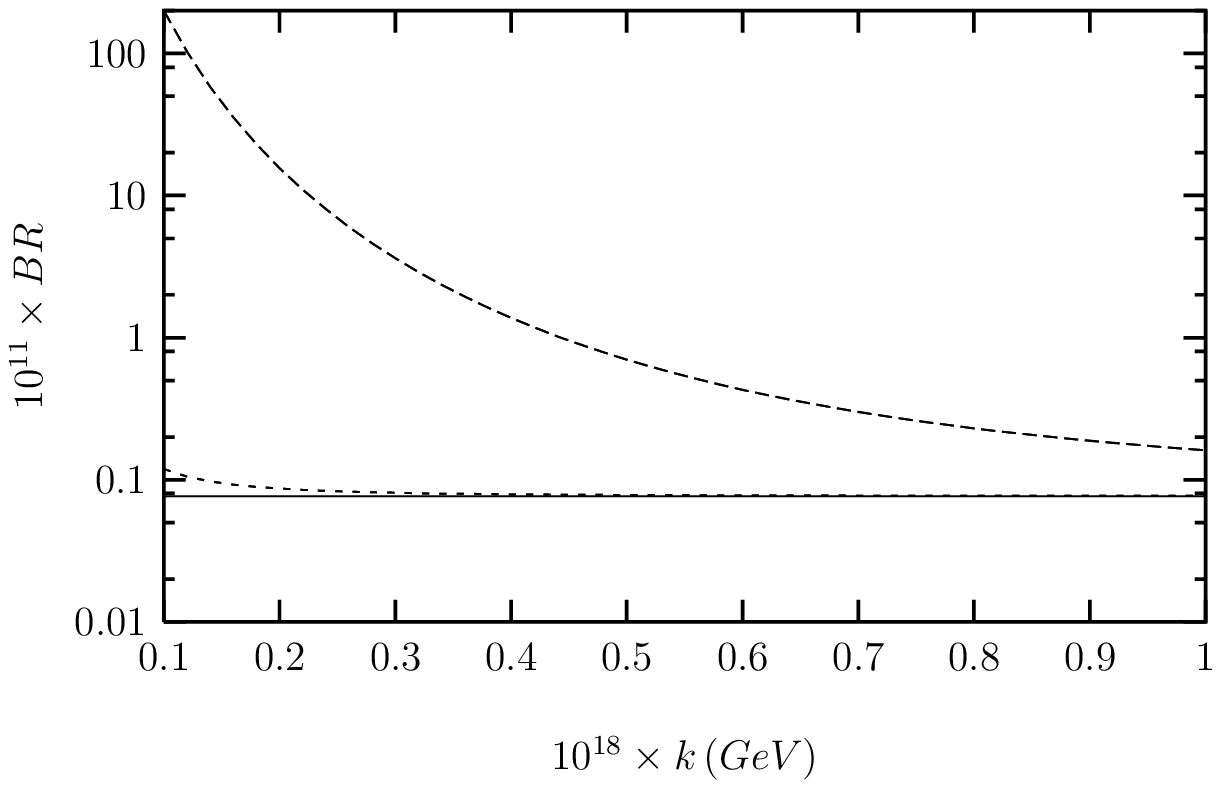} \vskip -3.0truein
\caption[]{The same as Fig.\ref{BRleptBulkmuegamk} but for
$\tau\rightarrow e \gamma$ decay and for $\bar{\xi}^{E}_{N,\tau
e}=0.1\, GeV$, $\bar{\xi}^{E}_{N,\tau\tau}=50\,GeV$.}
\label{BRleptBulktauegamk}
\end{figure}

\begin{figure}[htb]
\vskip -3.0truein \centering \epsfxsize=6.8in
\leavevmode\epsffile{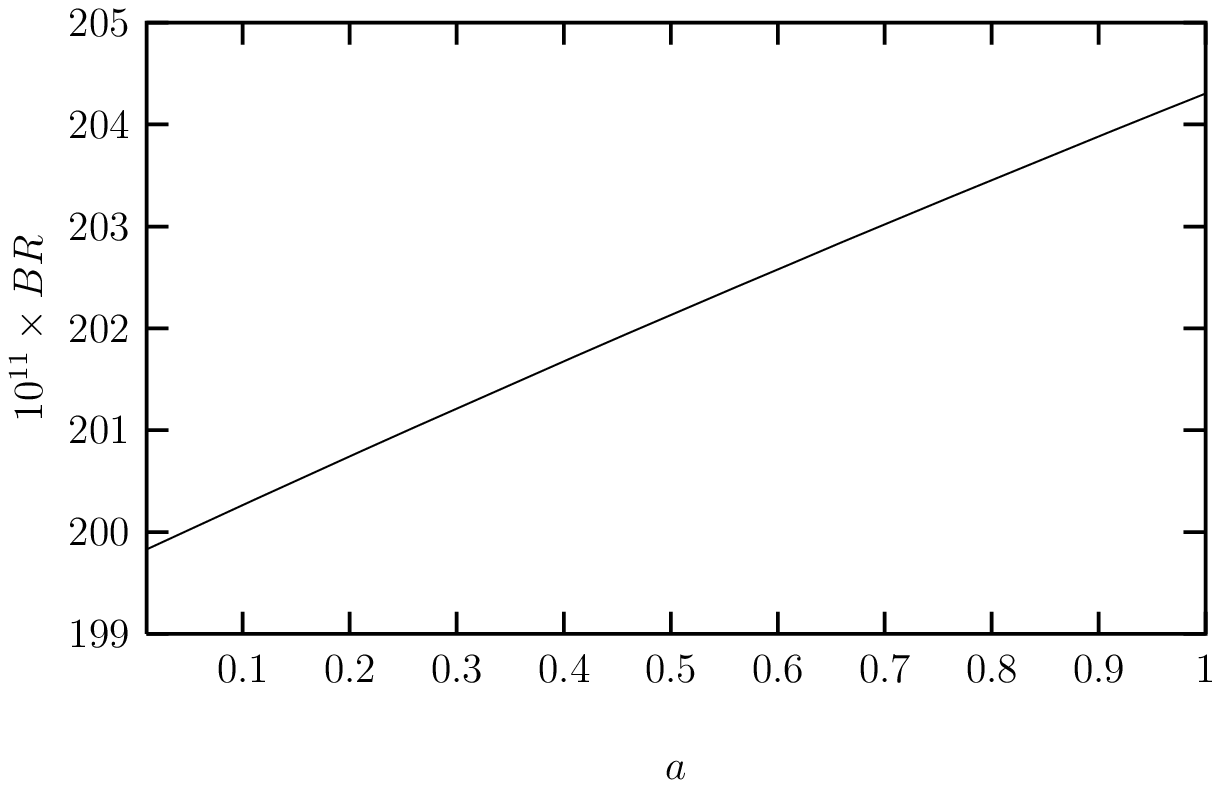} \vskip -3.0truein
\caption[]{ The same as Fig.\ref{BRleptBulkmuegamabehv} but for
$\tau\rightarrow e \gamma$ decay, for $\bar{\xi}^{E}_{N,\tau
e}=0.1\, GeV$, $\bar{\xi}^{E}_{N,\tau\tau}=50\,GeV$ and $k =
10^{17}\,GeV$.} \label{BRleptBulktauegamabehv}
\end{figure}

Fig.\ref{BRleptBulktauegamk}  is devoted to the parameter $k$
dependence of the BR of the LFV $\tau\rightarrow e \gamma$ decay
for $\bar{\xi}^{E}_{N,\tau e}=0.1\, GeV$,
$\bar{\xi}^{E}_{N,\tau\tau}=50\,GeV$. Here the solid (dashed,
short dashed) line represents the BR without KK modes of leptons
and new Higgs bosons (with KK modes of leptons and new Higgs
bosons for lepton location set II, set I), for $a=0.01$ and $0.1$.
This figure shows that the BR ($\tau\rightarrow e \gamma$) is of
the order of $10^{-12}$ without the internal lepton and new Higgs
boson KK mode contributions. The addition of these KK modes
results in that the BR enhances almost three orders for the small
values of the parameter $k$ and the lepton location set II. For
the set I, the BR enhances to the value almost two times larger
compared to the one without KK modes.
Fig.\ref{BRleptBulktauegamabehv} represents the parameter $a$
dependence of the BR ($\tau\rightarrow e \gamma$) for $k =
10^{17}\,GeV$, for the lepton location set II. It is observed that
the enhancement of the BR ($\tau\rightarrow e \gamma$) is greater
than $\sim 2.0\%$ in the range of $a$, $0.01 \leq a \leq 1.0$.
Similar to the previous decay, this is a small enhancement which
can not be determined.

\begin{figure}[htb]
\vskip -3.0truein \centering \epsfxsize=6.8in
\leavevmode\epsffile{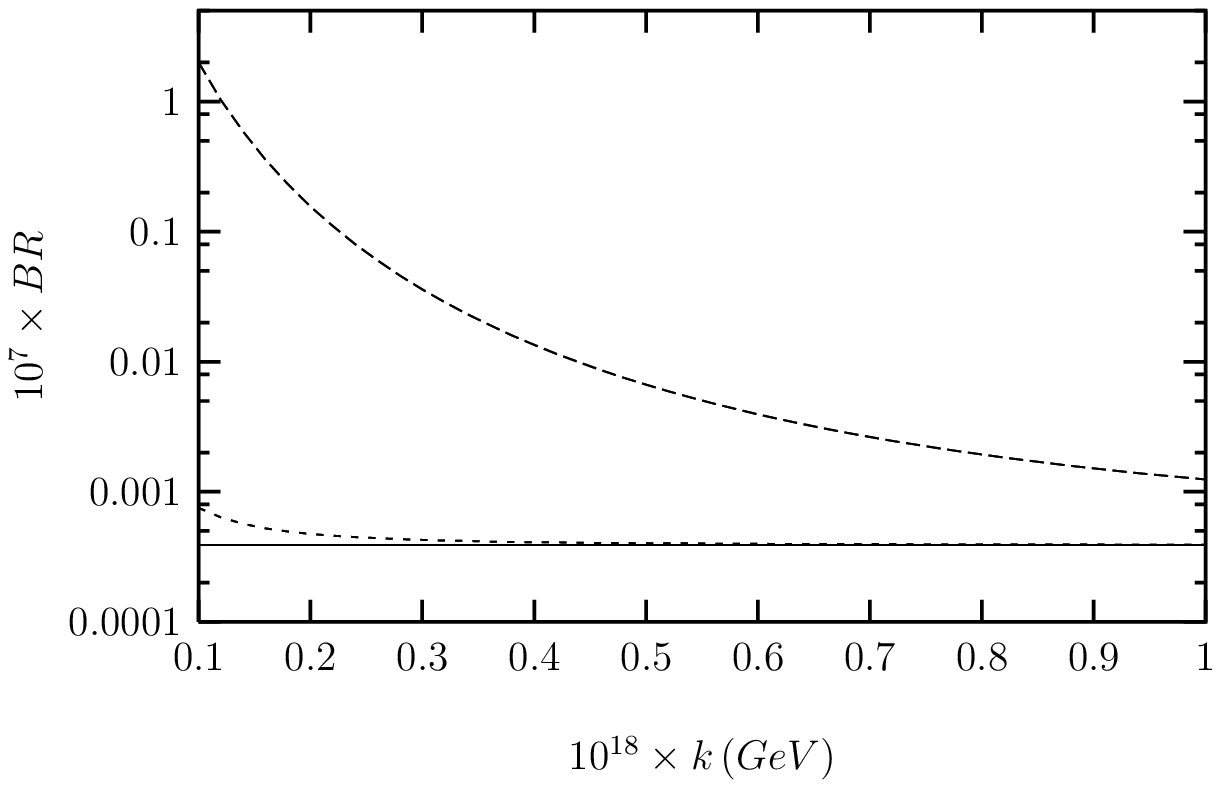} \vskip -3.0truein
\caption[]{The same as Fig.\ref{BRleptBulkmuegamk} but for
$\tau\rightarrow \mu \gamma$ decay and for $\bar{\xi}^{E}_{N,\tau
\mu}=1.0\, GeV$, $\bar{\xi}^{E}_{N,\tau\tau}=50\,GeV$.}
\label{BRleptBulktaumugamk}
\end{figure}

\begin{figure}[htb]
\vskip -3.0truein \centering \epsfxsize=6.8in
\leavevmode\epsffile{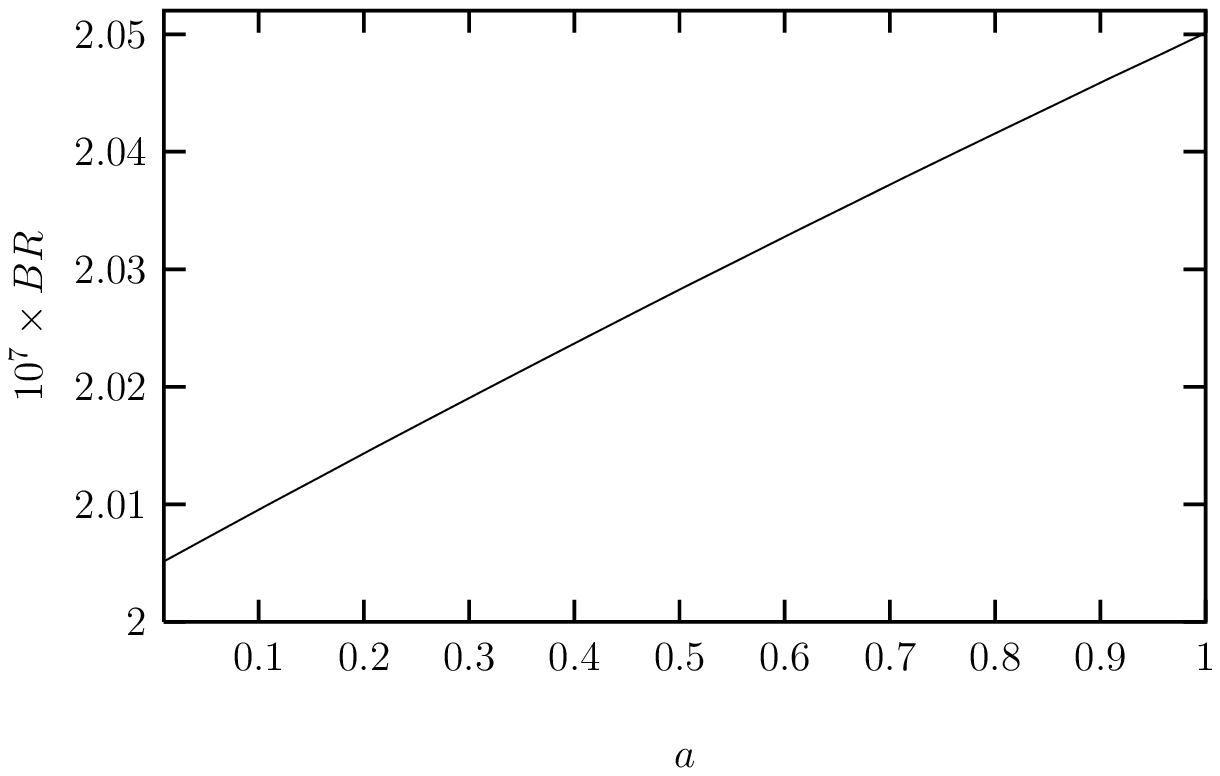} \vskip -3.0truein
\caption[]{The same as Fig.\ref{BRleptBulkmuegamabehv} but for
$\tau\rightarrow \mu \gamma$ decay, for $\bar{\xi}^{E}_{N,\tau
\mu}=1.0\, GeV$, $\bar{\xi}^{E}_{N,\tau\tau}=50\,GeV$ and $k =
10^{17}\,GeV$.} \label{BRleptBulktaumugamabehv}
\end{figure}

Fig.\ref{BRleptBulktaumugamk}  shows is the parameter $k$
dependence of the BR of the  LFV $\tau\rightarrow \mu \gamma$
decay for $\bar{\xi}^{E}_{N,\tau \mu}=1.0\, GeV$,
$\bar{\xi}^{E}_{N,\tau\tau}=50\,GeV$. Here the solid (dashed,
short dashed) line represents the BR without KK modes of leptons
and new Higgs bosons (with KK modes of leptons and new Higgs
bosons for lepton location set II, set I), for $a=0.01$ and $0.1$.
The BR ($\tau\rightarrow \mu \gamma$) is at the order of the
magnitude of $10^{-10}$ without the internal lepton and new Higgs
boson KK mode contributions. The lepton and Higgs boson KK modes
cause more than three order enhancement in the BR for the small
values of the parameter $k$ and for the lepton location set II.
For the set I, this enhancement is  almost two times of the BR
without KK modes. In Fig.\ref{BRleptBulktaumugamabehv} we present
the parameter $a$ dependence of the BR ($\tau\rightarrow \mu
\gamma$) for $k = 10^{17}\,GeV$ and the lepton location set II. We
observe that the enhancement of the BR ($\tau\rightarrow
\mu\gamma$) is more than $\sim 2.0\%$ in the range of $a$, $0.01
\leq a \leq 1.0$.

At this stage, we would like to summarize our results. For the
brane leptons, the contribution of the Higgs boson KK modes to the
BRs of the radiative LFV decays is too small to be detected.
However, if one considers that the leptons are also accessible to
the extra dimension, there exists a considerable enhancement in
the BRs, especially for the small values of the parameter $k$.
This enhancement occurs for the lepton location set II and it is
due to the fact that the left handed leptons (KK modes), which are
near to the visible brane, have enhanced couplings to the new
Higgs bosons, which are also localized near the visible brane.
Finally, we observe that, the BRs are weakly sensitive to the
parameter $a$ which regulates the localization of the new Higgs
doublet in the extra dimension. With the more accurate forthcoming
measurements of the BRs of the LFV decays it would be possible to
test the existence of the warped extra dimensions and, to get a
considerable information which fields are accessible to the extra
dimension. 
\newpage
\appendix

\vskip0.8cm \noindent \centerline{\Large \bf Appendix} \vskip0.4cm
\noindent
\section{The mass matrix of new Higgs boson}
The Higgs potential which creates the masses of neutral CP even
and CP odd Higgs bosons, $S=h^0$, $A^0$ reads
\begin{eqnarray}
V_{m_S}&=&c^\prime_h [Re(\phi_1^+ \phi_2)]^2 + c^\prime_A
[Im(\phi_1^+ \phi_2)]^2 \, , \label{masspotential}
\end{eqnarray}
where  $\phi_1$ and  $\phi_2$ are given in (eq. (\ref{choice})).
After the electrowek breaking the SM Higgs acquires a vacuum
expectation value (see (eq.(\ref{choice2})) and the mass
lagrangian of new CP even Higgs boson ($h^0$) \footnote{The
similar mass lagrangian appears for the CP odd Higgs boson $A^0$
with the replacement $c^\prime_h\rightarrow c^\prime_A$.} becomes
\begin{eqnarray}
{\cal{L}}_S=\frac{1}{2}\,\sum_{n=1}^\infty\, m_{n\,S}^2\,(S^{(n)}
(x))^2\,+c^\prime_h\,\frac{v^2}{2}\,\Bigg(S^{(0)}(x)+
\sum_{n=1}^\infty\,S^{(n)}(x) \,\alpha_n \Bigg)^2
 \label{massLagrangian} \, ,
\end{eqnarray}
with $c_h=c^\prime_h\,f^2_0(\pi\,R)$,
$\alpha_n=\frac{f_n(\pi\,R)}{f_0(\pi\,R)}$ and $m_S^2=c_h\,v^2$,
$S=h^0$. By using the mass lagrangian  the $S$ boson mass matrix
is obtained as (see \cite{MuckPilaftsisRuckl} and
\cite{iltanZl1l2Extr} for boson mass matrix in the one and two
non-universal extra dimensions, \cite{Batell} for $U(1)_Y$ gauge
boson mass matrix.):
\begin{equation}
M_{S}^2 \, = \, \left(
\begin{array}{cccc}
m_S^2 & \alpha_1 \, m_S^2 & \alpha_2 \, m_S^2 & \cdots \\
\alpha_1 \, m_S^2 & m_{1\,S}^2+ m_S^2\,\alpha_1^2 &
m_S^2\,\alpha_1\,
\alpha_2 & \cdots \\
\alpha_2 \, m_S^2& \,\alpha_2\,\alpha_1\, m_S^2 & m_{2\,S}^2+
m_S^2\,\alpha_2^2& \cdots \\\vdots & \vdots & \vdots & \ddots
\end{array}
\right) \quad ,
\end{equation}
and the determinant equation reads
\begin{equation}
\det \Big(M_{S}^2 - \lambda\,I \Big) = \Big(\prod_{n=1}^{\infty} (
m_{n\,S}^2 - \lambda )\Big)\, \Big( m_S^2 - \lambda- \lambda \,
m_S^2 \sum^{\infty}_{n=1} \, \frac{\alpha_n}{m_{n\,S}^2  -
\lambda}\Big) = 0 \, . \label{detone}
\end{equation}
This equation is used to calculate the physical masses of zero and
KK modes of $S$ bosons and their eigenstates. To leading order,
the physical $S$ boson mass (the zero mode one) reads
\begin{equation}
 (m^{phys}_S)^2 = m_S^2\,\Bigg( 1+\sum_{n=1}^\infty\,
 \frac{m_S^2\, \alpha_n^2}
 {m_{n\,S}^2} \Bigg) \label{transcone}\, ,
\end{equation}
since $m_{n\,S}>> m_S$. Notice that in our numerical calculations
we do not take into account the additional effects coming from the
mixing because they are suppressed due to the fact that the KK
mode masses are considerably larger compared to the zero mode one.

\section{The amplitudes appearing in the text}
Here, we present the amplitudes which appear in the calculation of
the decay widths of the radiative decays under consideration.

In the case that the charged leptons are restricted to the brane,
the amplitudes $A^0_1$, $A^0_2$, $A^{S\, KK}_1$ and $A^{S\, KK}_2$
in eq.(\ref{A1A1S}) read
\begin{eqnarray}
A^0_1&=&Q_{\tau} \frac{1}{4\,m_{\tau}^2} \Bigg \{ m_{l_1}\,
\bar{\xi}^{E}_{N, \tau l_2}\, \bar{\xi}^{E}_{N,\tau
l_1}\,\int_0^1\,dx \, \int_0^{1-x}\,dy\,\,
x\,(x+y-1)\,(\frac{z_{h^0}}{L_{h^0}}+\frac{z_{A^0}}{L_{A^0}})
\nonumber
\\ &-& m_{l_2}\, \bar{\xi}^{E}_{N,l_2 \tau }\,
\bar{\xi}^{E}_{N,l_1\tau}\, \int_0^1\,dx \,
\int_0^{1-x}\,dy\,x\,y\,(\frac{z_{h^0}}{L_{h^0}}+\frac{z_{A^0}}{L_{A^0}})
 \nonumber
\\ &+& m_{\tau}\,\bar{\xi}^{E}_{N,\tau l_2}\, \bar{\xi}^{E}_{N,l_1\tau}\,
\int_0^1\,dx \,
\int_0^{1-x}\,dy\,(x-1)\,(\frac{z_{h^0}}{L_{h^0}}-\frac{z_{A^0}}{L_{A^0}})
\Bigg\}
\nonumber \,\, , \\
A^0_2&=&Q_{\tau} \frac{1}{4\,m_{\tau}^2} \Bigg \{ -m_{l_1}\,
\bar{\xi}^{E}_{N, l_2 \tau }\, \bar{\xi}^{E}_{N,l_1 \tau
}\,\int_0^1\,dx \, \int_0^{1-x}\,dy\,\,
x\,(x+y-1)\,(\frac{z_{h^0}}{L_{h^0}}+\frac{z_{A^0}}{L_{A^0}})
\nonumber
\\ &+& m_{l_2}\, \bar{\xi}^{E}_{N,\tau l_2}\,
\bar{\xi}^{E}_{N, \tau l_1}\, \int_0^1\,dx \,
\int_0^{1-x}\,dy\,x\,y\,(\frac{z_{h^0}}{L_{h^0}}+\frac{z_{A^0}}{L_{A^0}})
 \nonumber
\\ &-& m_{\tau}\,\bar{\xi}^{E}_{N,l_2\tau }\, \bar{\xi}^{E}_{N,\tau l_1}\,
\int_0^1\,dx \,
\int_0^{1-x}\,dy\,(x-1)\,(\frac{z_{h^0}}{L_{h^0}}-\frac{z_{A^0}}{L_{A^0}})
\Bigg\}
 \,\, , \label{A1A20}
\end{eqnarray}
\begin{eqnarray}
A^{S\, KK}_1&=&Q_{\tau} \frac{1}{4\,m_{\tau}^2}\,
\sum_{n=1}^\infty \,\Bigg \{ m_{l_1}\, \bar{\xi}^{E\,n}_{N,\tau
l_2}\, \bar{\xi}^{E\,n}_{N,\tau l_1}\,\int_0^1\,dx \,
\int_0^{1-x}\,dy\,\,
x\,(x+y-1)\,(\frac{z_{h^0\,n}}{L_{h^0\,n}}+\frac{z_{A^0\,n}}{L_{A^0\,n}})
\nonumber
\\ &-& m_{l_2}\, \bar{\xi}^{E\,n}_{N, l_2 \tau }\,
\bar{\xi}^{E\,n}_{N,l_1\tau}\, \int_0^1\,dx \,
\int_0^{1-x}\,dy\,x\,y\,(\frac{z_{h^0\,n}}{L_{h^0\,n}}+\frac{z_{A^0\,n}}
{L_{A^0\,n}})
 \nonumber
\\ &+& m_{\tau}\,\bar{\xi}^{E\,n}_{N, \tau l_2}\,
\bar{\xi}^{E\,n}_{N,l_1\tau}\, \int_0^1\,dx \,
\int_0^{1-x}\,dy\,(x-1)\,(\frac{z_{h^0\,n}}{L_{h^0\,n}}-\frac{z_{A^0\,n}}
{L_{A^0\,n}}) \Bigg\}
\nonumber \,\, , \\
A^{S\, KK}_2&=&Q_{\tau} \frac{1}{4\,m_{\tau}^2} \,
\sum_{n=1}^\infty \,\Bigg \{ -m_{l_1}\, \bar{\xi}^{E\,n}_{N, l_2
\tau }\, \bar{\xi}^{E\,n}_{N, l_1 \tau}\,\int_0^1\,dx \,
\int_0^{1-x}\,dy\,\,
x\,(x+y-1)\,(\frac{z_{h^0\,n}}{L_{h^0\,n}}+\frac{z_{A^0\,n}}{L_{A^0\,n}})
\nonumber
\\ &+& m_{l_2}\, \bar{\xi}^{E\,n}_{N, \tau l_2 }\,
\bar{\xi}^{E\,n}_{N,\tau l_1}\, \int_0^1\,dx \,
\int_0^{1-x}\,dy\,x\,y\,(\frac{z_{h^0\,n}}{L_{h^0\,n}}+\frac{z_{A^0\,n}}
{L_{A^0\,n}})
 \nonumber
\\ &-& m_{\tau}\,\bar{\xi}^{E\,n}_{N,l_2\tau }\,
\bar{\xi}^{E\,n}_{N,\tau l_1}\,\int_0^1\,dx \,
\int_0^{1-x}\,dy\,(x-1)\,(\frac{z_{h^0\,n}}{L_{h^0\,n}}-\frac{z_{A^0\,n}}
{L_{A^0\,n}}) \Bigg\}
 \,\, , \label{A1A2SKK}
\end{eqnarray}
where
\begin{eqnarray}
L_{S}&=&z_S+x^2\,z_S+x\Big( 1+(y-2)\,z_S \Big)\, ,\nonumber
\\
L_{S\,n}&=&z_{S\,n}+x^2\,z_{S\,n}+x\Big( 1+(y-2)\,z_{S\,n} \Big)
\,\, , \label{LS}
\end{eqnarray}
with  $z_{S}=\frac{m^2_{\tau}}{m^2_{S}}$,
$z_{S\,n}=\frac{m^2_{\tau}}{m^2_{n\,S}}$. Here ,
$l_1\,(l_2)=\tau;\mu\,(\mu$ or $e; e)$, $c_1=\frac{G_F^2
\alpha_{em} m^3_{l_1}}{32 \pi^4}$, $A_1$ ($A_2$) is the left
(right) chiral amplitude, $Q_{\tau}$ is the charge of tau lepton
and $m_{n\,S}$ is the internal Higgs KK mode mass (see
eq.(\ref{mn})). Notice that we take the Yukawa couplings real.

If the charged leptons are also accessible to the extra dimension
the amplitudes $A^0_{1\,(2)}$, $A^{S\, KK}_{1\,(2)}$ and $A^{S,\,
l\, KK}_{1\,(2)}$ (see eq.(\ref{A120SlKK})) are
\begin{eqnarray}
A^0_1&=&Q_{\tau} \frac{1}{4\,m_{\tau}^2} \Bigg \{ m_{l_1}\,
\bar{\xi}^{E}_{N, \tau l_2}\, \bar{\xi}^{E}_{N,\tau
l_1}\,\int_0^1\,dx \, \int_0^{1-x}\,dy\,\,
x\,(x+y-1)\,(\frac{z_{h^0}}{L_{h^0}}+\frac{z_{A^0}}{L_{A^0}})
\nonumber
\\ &-& m_{l_2}\, \bar{\xi}^{E}_{N,l_2 \tau }\,
\bar{\xi}^{E}_{N,l_1\tau}\, \int_0^1\,dx \,
\int_0^{1-x}\,dy\,x\,y\,(\frac{z_{h^0}}{L_{h^0}}+\frac{z_{A^0}}{L_{A^0}})
 \nonumber
\\ &+& m_{\tau}\,\bar{\xi}^{E}_{N,\tau l_2}\, \bar{\xi}^{E}_{N,l_1\tau}\,
\int_0^1\,dx \,
\int_0^{1-x}\,dy\,(x-1)\,(\frac{z_{h^0}}{L_{h^0}}-\frac{z_{A^0}}{L_{A^0}})
\Bigg\}
\nonumber \,\, , \\
A^0_2&=&Q_{\tau} \frac{1}{4\,m_{\tau}^2} \Bigg \{ -m_{l_1}\,
\bar{\xi}^{E}_{N, l_2 \tau }\, \bar{\xi}^{E}_{N,l_1 \tau
}\,\int_0^1\,dx \, \int_0^{1-x}\,dy\,\,
x\,(x+y-1)\,(\frac{z_{h^0}}{L_{h^0}}+\frac{z_{A^0}}{L_{A^0}})
\nonumber
\\ &+& m_{l_2}\, \bar{\xi}^{E}_{N,\tau l_2}\,
\bar{\xi}^{E}_{N, \tau l_1}\, \int_0^1\,dx \,
\int_0^{1-x}\,dy\,x\,y\,(\frac{z_{h^0}}{L_{h^0}}+\frac{z_{A^0}}{L_{A^0}})
 \nonumber
\\ &-& m_{\tau}\,\bar{\xi}^{E}_{N,l_2\tau }\, \bar{\xi}^{E}_{N,\tau l_1}\,
\int_0^1\,dx \,
\int_0^{1-x}\,dy\,(x-1)\,(\frac{z_{h^0}}{L_{h^0}}-\frac{z_{A^0}}{L_{A^0}})
\Bigg\}
 \,\, , \label{A1A220}
\end{eqnarray}
\begin{eqnarray}
A^{S\, KK}_1&=&Q_{\tau} \frac{1}{4\,m_{\tau}^2}\,
\sum_{n=1}^\infty \,\Bigg \{ m_{l_1}\, \bar{\xi}^{E\,n0}_{N,\tau
l_2}\, \bar{\xi}^{E\,n0}_{N,\tau l_1}\,\int_0^1\,dx \,
\int_0^{1-x}\,dy\,\,
x\,(x+y-1)\,(\frac{z_{h^0\,n}}{L_{h^0\,n}}+\frac{z_{A^0\,n}}{L_{A^0\,n}})
\nonumber
\\ &-& m_{l_2}\, \bar{\xi}^{E\,n0}_{N, l_2 \tau }\,
\bar{\xi}^{E\,n0}_{N,l_1\tau}\, \int_0^1\,dx \,
\int_0^{1-x}\,dy\,x\,y\,(\frac{z_{h^0\,n}}{L_{h^0\,n}}+\frac{z_{A^0\,n}}
{L_{A^0\,n}})
 \nonumber
\\ &+& m_{\tau}\,\bar{\xi}^{E\,n0}_{N, \tau l_2}\,
\bar{\xi}^{E\,n0}_{N,l_1\tau}\, \int_0^1\,dx \,
\int_0^{1-x}\,dy\,(x-1)\,(\frac{z_{h^0\,n}}{L_{h^0\,n}}-\frac{z_{A^0\,n}}
{L_{A^0\,n}}) \Bigg\}
\nonumber \,\, , \\
A^{S\, KK}_2&=&Q_{\tau} \frac{1}{4\,m_{\tau}^2} \,
\sum_{n=1}^\infty \,\Bigg \{ -m_{l_1}\, \bar{\xi}^{E\,n0}_{N, l_2
\tau }\, \bar{\xi}^{E\,n0}_{N, l_1 \tau}\,\int_0^1\,dx \,
\int_0^{1-x}\,dy\,\,
x\,(x+y-1)\,(\frac{z_{h^0\,n}}{L_{h^0\,n}}+\frac{z_{A^0\,n}}{L_{A^0\,n}})
\nonumber
\\ &+& m_{l_2}\, \bar{\xi}^{E\,n0}_{N, \tau l_2 }\,
\bar{\xi}^{E\,n0}_{N,\tau l_1}\, \int_0^1\,dx \,
\int_0^{1-x}\,dy\,x\,y\,(\frac{z_{h^0\,n}}{L_{h^0\,n}}+\frac{z_{A^0\,n}}
{L_{A^0\,n}})
 \nonumber
\\ &-& m_{\tau}\,\bar{\xi}^{E\,n0}_{N,l_2\tau }\,
\bar{\xi}^{E\,n0}_{N,\tau l_1}\,\int_0^1\,dx \,
\int_0^{1-x}\,dy\,(x-1)\,(\frac{z_{h^0\,n}}{L_{h^0\,n}}-\frac{z_{A^0\,n}}
{L_{A^0\,n}}) \Bigg\}
 \,\, , \label{A1A22SKK}
\end{eqnarray}
\begin{eqnarray}
A^{l\, KK}_1&=&
\frac{Q_{\tau}}{48\,m_{\tau}^2}\,\sum_{n=1}^{\infty} \, \Bigg \{
\frac{m_{\tau}^2}{m^2_{nR}} \,\,m_{l_1}\,(\bar{\xi}^{E\,0n}_{N,
l_2\tau})^\dagger\, (\bar{\xi}^{E\,0n}_{N,l_1\tau})^\dagger\,
\Big(G (z_{nR, h^0})+G (z_{nR, A^0})\Big)\nonumber \\&+&
\frac{m_{\tau}^2}{m^2_{nL}} \,\,m_{l_2}\,\bar{\xi}^{E\,0n}_{N,
l_2\tau}\, \bar{\xi}^{E\,0n}_{N,l_1\tau}\, \Big(G (z_{nL, h^0})+G
(z_{nL, A^0})\Big) \Bigg\}
\nonumber \,\, , \\
A^{l\, KK}_2&=&
\frac{-Q_{\tau}}{48\,m_{\tau}^2}\,\sum_{n=1}^{\infty} \, \Bigg \{
\frac{m_{\tau}^2}{m^2_{nR}} \,m_{l_2}\,(\bar{\xi}^{E\,0n}_{N,
l_2\tau})^\dagger\, (\bar{\xi}^{E\,0n}_{N,l_1\tau})^\dagger\,
\Big(G (z_{nR, h^0})+G (z_{nR, A^0})\Big)\nonumber \\&+&
\frac{m_{\tau}^2}{m^2_{nL}}\,m_{l_1} \,\bar{\xi}^{E\,0n}_{N,
l_2\tau}\, \bar{\xi}^{E\,0n}_{N,l_1\tau}\, \Big(G (z_{nL, h^0})+G
(z_{nL, A^0})\Big) \Bigg\}
 \,\, , \label{A1A22n}
\end{eqnarray}
\begin{eqnarray}
A^{S, l\, KK}_1&=&
\frac{Q_{\tau}}{48\,m_{\tau}^2}\,\sum_{n,m=1}^{\infty} \, \Bigg \{
\frac{m_{\tau}^2}{m^2_{nR}} \,\,m_{l_1}\,(\bar{\xi}^{E\,mn}_{N,
l_2\tau})^\dagger\, (\bar{\xi}^{E\,mn}_{N,l_1\tau})^\dagger\,
\Big(G (z_{nR, h^0\,m})+G (z_{nR, A^0\,m})\Big)\nonumber \\&+&
\frac{m_{\tau}^2}{m^2_{nL}} \,\,m_{l_2}\,\bar{\xi}^{E\,mn}_{N,
l_2\tau}\, \bar{\xi}^{E\,mn}_{N,l_1\tau}\, \Big(G (z_{nL,
h^0\,m})+G (z_{nL, A^0\,m})\Big) \Bigg\}
\nonumber \,\, , \\
A^{S, l\, KK}_2&=&
\frac{-Q_{\tau}}{48\,m_{\tau}^2}\,\sum_{n,m=1}^{\infty} \, \Bigg
\{ \frac{m_{\tau}^2}{m^2_{nR}} \,m_{l_2}\,(\bar{\xi}^{E\,mn}_{N,
l_2\tau})^\dagger\, (\bar{\xi}^{E\,mn}_{N,l_1\tau})^\dagger\,
\Big(G (z_{nR, h^0\,m})+G (z_{nR, A^0\,m})\Big)\nonumber \\&+&
\frac{m_{\tau}^2}{m^2_{nL}}\,m_{l_1} \,\bar{\xi}^{E\,mn}_{N,
l_2\tau}\, \bar{\xi}^{E\,mn}_{N,l_1\tau}\, \Big(G (z_{nL,
h^0\,m})+G (z_{nL, A^0\,m})\Big) \Bigg\}
 \, . \label{A1A22mn}
\end{eqnarray}
Here $A_1$ ($A_2$) is the left (right) chiral amplitude,
$l_1\,(l_2)=\tau;\mu\,(\mu$ or $e; e)$, the functions $F (w)$, $G
(w)$ are
\begin{eqnarray}
F (w)&=&\frac{w\,(3-4\,w+w^2+2\,ln\,w)}{(1-w)^3} \, , \nonumber \\
G (w)&=&-\frac{w\,(2+3\,w-6\,w^2+w^3+ 6\,w\,ln\,w)}{(1-w)^4} \,\,
, \label{functions2}
\end{eqnarray}
$c_1=\frac{G_F^2 \alpha_{em} m^3_{l_1}}{32 \pi^4}$, ,
$z_{S}=\frac{m^2_{\tau}}{m^2_{S}}$,
$z_{S\,n}=\frac{m^2_{\tau}}{m^2_{n\,S}}$, $z_{nL(nR),
S}=\frac{m^2_{nL\,(nR)}}{m^2_{S}}$, $z_{nL(nR),
S\,m}=\frac{m^2_{nL\,(nR)}}{m^2_{m\,S}}$ with left (right) handed
internal lepton KK mode mass $m_{nL\,(nR)}$ (eq.(\ref{mnLR})). In
eqs. (\ref{A1A220}) and (\ref{A1A22SKK}) the functions $L_{S}$ and
$L_{S\,n}$ are given in (eq.(\ref{LS})). In the case that the
incoming and outgoing lepton masses are ignored in  the functions
$L_{S}$ and $L_{S\,n}$ one gets the integrated form of
$A^0_{1(2)}$ as
\begin{eqnarray}
A^0_1&=&Q_{\tau} \frac{1}{48\,m_{\tau}^2} \Bigg \{ 6\,m_\tau\,
\bar{\xi}^{E}_{N,\tau l_2 }\, \bar{\xi}^{E}_{N,l_1 \tau }\, \Big(
F (z_{h^0})-F (z_{A^0}) \Big)+ m_{l_1}\, \bar{\xi}^{E}_{N,\tau
l_2}\, \bar{\xi}^{E}_{N, \tau l_1}\, \Big(G
(z_{h^0})+G(z_{A^0})\Big) \Bigg\}
\nonumber \,\, , \\
A^0_2&=&-Q_{\tau} \frac{1}{48\,m_{\tau}^2} \Bigg \{6\,m_\tau\,
\bar{\xi}^{E}_{N, l_2 \tau}\, \bar{\xi}^{E}_{N,\tau l_1}\, \Big(F
(z_{h^0})-F(z_{A^0})\Big)+ m_{l_1}\, \bar{\xi}^{E}_{N,l_2\tau}\,
\bar{\xi}^{E}_{N,l_1 \tau}\, \Big( G (z_{h^0})+G (z_{A^0})\Big)
\Bigg\}
 \, . \nonumber \\ \label{A1A2Intform}
\end{eqnarray}
Notice that, for the amplitudes $A^{l\, KK}_{1(2)}$ and $A^{S, l\,
KK}_{1(2)}$, the incoming and outgoing lepton masses are ignored
in the functions $L_{S}$ and $L_{S\,n}$  since the internal KK
leptons are heavy.
\section{The construction of zero mode and KK mode leptons}
This Appendix is devoted to the construction of the zero mode and
KK mode leptons in the case that the leptons are localized in the
extra dimension with the help of the Dirac mass term given in
(eq.(\ref{massterm})). We start with the expansion of the bulk
fermion  as
\begin{eqnarray}
\psi(x^\mu,y)=\frac{1}{\sqrt{2\,\pi\,R}}\,\sum_{n=0}^\infty\,
\psi^{(n)}(x^\mu)\, e^{2\,\sigma}\, \chi_n(y) \label{psiKK} \, .
\end{eqnarray}
By using the normalization
\begin{eqnarray}
\frac{1}{2\,\pi\,R}\,\int_{-\pi\,R}^{\pi\,R}\,dy\,e^\sigma \,
\chi_n(y)\,\chi_m(y)=\delta_{nm} \label{norm} \, ,
\end{eqnarray}
and the Dirac equation the zero mode fermion is obtained as
\begin{eqnarray}
\chi_0(y)=N_0\, e^{-r\,\sigma}\label{0mode} \, ,
\end{eqnarray}
where $r=m/k$ and $N_0$ is the normalization constant:
\begin{eqnarray}
N_0=\sqrt{\frac{k\,\pi \,R\,(1-2\,r)}{e^{k\,\pi
\,R\,(1-2\,r)}-1}}\label{norm0mode} \, .
\end{eqnarray}
The appropriately normalized solution
\begin{eqnarray}
\chi'_0(y)=e^{-\frac{\sigma}{2}}\,\chi_0(y) \label{0modep}
\end{eqnarray}
is localized in the extra dimension  where the localization is
regulated by the parameter $r$. For $r>\frac{1}{2}$
($r<\frac{1}{2}$) this solution is localized  near the hidden
(visible) brane and it has a constant profile for $r=\frac{1}{2}$.

Now, we are ready construct the SM leptons. What we need is  to
consider $SU(2)_L$ doublet $\psi_L$ and singlet $\psi_R$ with
separate $Z_2$ projection conditions: $Z_2\psi_R=\gamma_5 \psi_R$
and $Z_2\psi_L=-\gamma_5 \psi_L$ (see for example \cite{Hisano}).
Finally we get the leptons accessible to the extra dimension as
\begin{eqnarray}
l_{i L}(x^\mu,y)&=&\frac{1}{\sqrt{2\,\pi\,R}}\, e^{2\,\sigma}\,
l_{i L}^{(0)}(x^\mu)\, \chi_{i\,L 0}(y)\nonumber
\\ &+& \frac{1}{\sqrt{2\,\pi\,R}}\,\sum_{n=1}^\infty\, e^{2\,\sigma}\,
\Bigg( l_{i L}^{(n)}(x^\mu)\, \chi^{l}_{i\,L n}(y)+l_{i
R}^{(n)}(x^\mu)\, \chi^{l}_{i\,R n}(y)\Bigg)\, , \nonumber
\\
E_{j R}(x^\mu,y)&=& \frac{1}{\sqrt{2\,\pi\,R}}\,
e^{2\,\sigma}\,E_{j R}^{(0)}(x^\mu)\, \chi_{j\,R 0}(y)\nonumber
\\ &+&\frac{1}{\sqrt{2\,\pi\,R}}\,\sum_{n=0}^\infty\,
e^{2\,\sigma}\,\Bigg( E_{j R}^{(n)}(x^\mu)\, \chi^E_{j\,R n}(y)+
E_{j L}^{(n)}(x^\mu)\, \chi^E_{j\,L n}(y)\Bigg) \label{leptonKK}
\, .
\end{eqnarray}
Here the zero mode leptons $\chi_{i\,L 0}(y)$ and  $\chi_{j\,R
0}(y)$ are given in eq.(\ref{0mode}) with the replacements
$r\rightarrow r_{iL}$ and $r\rightarrow r_{jR}$, respectively.

The zero mode fermions can get mass through the $Z_2$ invariant
left handed fermion-right handed fermion-Higgs interaction,
$\bar{\psi}_R\,\psi_L\, H$\footnote{Here, we consider different
location parameters $r$ for each left handed and right handed part
of different flavors. The location parameters for fermion fields
are chosen so that this interaction creates the current masses of
fermions.}. If the SM Higgs field lives on the visible brane as in
our choice, the masses of fermions are calculated by using the
integral
\begin{eqnarray}
m_i=\frac{1}{2\,\pi\,R}\,\int_{-\pi\,R}^{\pi\,R}\,dy\,\lambda_5\,
\chi_{iL0}(y)\,\chi_{iR0}(y)\,<H>\,\delta(y-\pi\, R) \label{mi} \,
,
\end{eqnarray}
where $\lambda_5$ is the coupling in five dimensions and it can be
parametrized  in terms of the one in four dimensions, the
dimensionless coupling $\lambda$, $\lambda_5=\lambda/\sqrt{k}$.
Here the expectation value of the Higgs field $<H>$ reads
$<H>=v/\sqrt{k}$ where $v$ is the vacuum expectation
value\footnote{We take $v=0.043\,M_{Pl}$ to provide the measured
gauge boson masses \cite{Huber2} and choose $k\,R=10.83$ in order
to get the correct effective scale on the visible brane, i.e.,
$M_W=e^{-\pi\,k\,R}\, M_{pl}$ is of the order of TeV.}. Now, we
choose two different sets of location of charged lepton fields in
order to obtain the masses of different flavors.
\begin{table}[h]
        \begin{center}
\begin{tabular}{|c|c|c|c|c|}
  \hline
    & SET I  & SET  II  \\
  \hline \hline
& $r_L$ \,\,\,\,\, $r_R$  & \,$r_L$ \,\,\,\,\,  $r_R$ \\
\hline\hline
  e & 0.6710  \,\, 0.6710  & -0.4900 \,\, 0.8800  \\ \hline
  $\mu$  & 0.5826 \,\,  0.5826  & -0.4900 \,\, 0.7160  \\ \hline
  $\tau$ & 0.5273 \,\, 0.5273 & -0.4900 \,\,0.6249  \\ \hline
  \hline
\end{tabular}
\end{center}
\caption{Two possible locations of charged lepton fields. Here
$r_L$ and $r_R$ are left handed and right handed lepton field
location parameters, respectively.} \label{set}
\end{table}
In Set I, the left and right handed fields of the same flavor have
the same location, however, in set II, we choose the left handed
charged lepton locations the same for each flavor. For both cases
we estimate the left-right handed charged lepton locations by
respecting their current masses.

The $Z_2$ the projection condition $Z_2\psi=-\gamma_5 \psi$, used
in constructing the left handed fields on the branes, results in
that the left handed zero mode appears, the left (right) handed KK
modes appear (disappear) on the branes, with the boundary
conditions due to the Dirac mass term in the action
eq.(\ref{massterm}):
\begin{eqnarray}
\Big(\frac{d}{dy}-m \Big)\,\chi^l_{iLn}(y_0)=0 \, ,\nonumber \\
\chi^l_{iRn}(y_0)=0 \, , \label{nLbound}
\end{eqnarray}
where $y_0=0$ or $\pi\,R$. Using the Dirac equation for KK mode
leptons one gets the left handed lepton $\chi^l_{i\,L n}(y)$ that
lives on the visible brane as
\begin{eqnarray}
\chi^l_{iLn}(y)=N_{Ln}\, e^{\sigma/2} \Bigg(
J_{\frac{1}{2}-r_{iL}} (e^{\sigma}\,x_{nL})+c_L\,
Y_{\frac{1}{2}-r_{iL}} (e^{\sigma}\,x_{nL})\Bigg)\label{nLmode} \,
,
\end{eqnarray}
with the constant
\begin{eqnarray}
c_L=-\frac{J_{-r_{iL}-\frac{1}{2}}
(x_{nL})}{Y_{-r_{iL}-\frac{1}{2}} (x_{nL})} \, .  \label{cL}
\end{eqnarray}
Here, $N_{Ln}$ is the normalization constant and
$x_{nL}=\frac{m_{Ln}}{k}$. The functions $J_\beta(w)$ and
$Y_\beta(w)$ appearing in eq.(\ref{nLmode}) are the Bessel
function of the first kind and of the second kind, respectively.
On the other hand, the $Z_2$ projection condition
$Z_2\psi=\gamma_5 \psi$ is used in order to construct the right
handed fields on the branes and this ensures that the right handed
zero mode appears, the right (left) handed KK modes appear
(disappear) on the branes with the boundary conditions:
\begin{eqnarray}
\Big(\frac{d}{dy}+m \Big)\,\chi^E_{iRn}(y_0)=0 \,, \nonumber \\
\chi^E_{iLn}(y_0)=0 \, . \label{nRbound}
\end{eqnarray}
Again, using the Dirac equation for KK mode leptons, one gets the
right handed lepton $\chi^E_{i\,R n}(y)$ that lives on the visible
brane as
\begin{eqnarray}
\chi^E_{iRn}(y)=N_{Rn}\, e^{\sigma/2} \Bigg(
J_{\frac{1}{2}+r_{iR}} (e^{\sigma}\,x_{nR})+c_R\,
Y_{\frac{1}{2}+r_{iR}} (e^{\sigma}\,x_{nR})\Bigg)\label{nRmode} \,
,
\end{eqnarray}
with
\begin{eqnarray}
c_R=-\frac{J_{r_{iR}-\frac{1}{2}} (x_{nR})}{Y_{r_{iR}-\frac{1}{2}}
(x_{nR})} \, , \label{cR}
\end{eqnarray}
where $N_{Rn}$ is the normalization constant and
$x_{nR}=\frac{m_{Rn}}{k}$. Notice that the constant $c_L$, the
$n^{th}$ KK mode mass $m_{Ln}$ in eq.(\ref{nLmode}) and the
constant $c_R$, the $n^{th}$ KK mode mass $m_{Rn}$ in
eq.(\ref{nRmode}) are obtained by using the boundary conditions
eq.(\ref{nLbound}) and eq.(\ref{nRbound}), respectively. For
$m_{L(R)n}\ll k$ and $kR\gg 1$ they are approximated as:
\begin{eqnarray}
m_{Ln}&\simeq&
k\,\pi\,\Big(n-\frac{\frac{1}{2}-r}{2}+\frac{1}{4}\Big)\,
e^{-\pi\,k\,R} \nonumber \, ,\\
m_{Rn} &\simeq&
k\,\pi\,\Big(n-\frac{\frac{1}{2}+r}{2}+\frac{1}{4}\Big)\,e^{-\pi\,k\,R}
\,\,\,\,\,\,\,\, \mbox{for $r<0.5$} \nonumber \, ,\\
m_{Rn}&\simeq&
k\,\pi\,\Big(n+\frac{\frac{1}{2}+r}{2}-\frac{3}{4}\Big)\,e^{-\pi\,k\,R}
\,\,\,\,\,\,\,\, \mbox{for $r>0.5$}\label{mnLR} \, .
\end{eqnarray}
\newpage
\newpage
%
%

\begin{thebibliography}{1}
%
\bibitem{Brooks} M. L. Brooks et. al., MEGA Collaboration,
{\it Phys. Rev. Lett.} {\bf 83} 1521, (1999).
%
\bibitem{Hayasaka} K. Hayasaka et al.., {\it Phys.Lett.} {\bf D63} 20, (2005).
%
\bibitem{Ahmed} S. Ahmed et.al., CLEO Collaboration,
{\it Phys. Rev.} {\bf D61} 071101, (2000).
%
\bibitem{Roney} J.M. Roney and the BABAR Collaboration,
{\it Nucl. Phys. Proc. Suppl.} {\bf 144} 155, (2005).
%
\bibitem{Aubert} B. Aubert et. al., BABAR Collaboration,
SLAC-PUB-11028, BABAR-PUB-04-049, Feb. 2005, 7. pp, {\it Phys.
Rev. Lett.} {\bf 95} 041802, (2005).
%
\bibitem{Nicolo} Donato Nicolo, MUEGAMMA Collaboration,
{\it Nucl. Instrum. Meth} {\bf A503} 287, (2003).
%
\bibitem{Yamada} S. Yamada, {\it Nucl. Phys. Proc. Suppl.} {\bf 144} 185,
(2005).
%
\bibitem{Lee} T. D. Lee, {\it Phys. Rev.} {\bf D8} 1226, (1973).
%
\bibitem{Glashow} S. Glashow and S. Weinberg,
{\it Phys. Rev.} {\bf D15} 1958,  (1977).
%
\bibitem{Gunion} For a review see J. Gunion, H. Haber, G. Kane, and
S. Dawson, The Higgs Hunter's Guide ~Addison-Wesley, New York,
(1990).
%
\bibitem{Atwood} D. Atwood, L. Reina and A. Soni,
{\it Phys. Rev.} {\bf D55} 3156, (1997).
%
\bibitem{Iltan1} E. O. Iltan, {\it Phys. Rev.} {\bf D64} 115005, (2001).
%
\bibitem{Iltan11} E. O. Iltan,{\it Phys. Rev.} {\bf D64} 013013, (2001)
%
\bibitem{Diaz} R. Diaz, R. Martinez and J. A. Rodriguez,
{\it Phys. Rev.} {\bf D63} 095007, (2001).
%
\bibitem{IltanExtrDim} E. O. Iltan, {\it JHEP} {\bf 0402} 20, (2004).
%
%
%
%
\bibitem{IltanLFVRS} E. O. Iltan, {\it Int. J. Mod. Phys.} {\bf
A23} 1055, (2008).
%
\bibitem{Diaz2} R. Diaz,  R. Martinez, J. A. Rodriguez,
{\it Phys. Rev.} {\bf D67} 075011, (2003).
%
\bibitem{Barbieri1} R. Barbieri and L. J. Hall,
{\it Phys. Lett.} {\bf B338} 212, (1994).
%
\bibitem{Barbieri2} R. Barbieri, L. J. Hall and A. Strumia, {\it Nucl. Phys.}
{\bf B445} 219, (1995).
%
\bibitem{Barbieri3} R. Barbieri, L. J. Hall and A. Strumia,
{\it Nucl. Phys.} {\bf B449} 437, (1995).
%
\bibitem{Barbieri4} P. Ciafaloni, A. Romanino and A. Strumia,
IFUP-YH-42-95.
%
\bibitem{Barbieri5} T. V. Duong, B. Dutta and E. Keith, {\it Phys. Lett.}
{\bf B378} 128, (1996).
%
\bibitem{Barbieri6} G. Couture, et. al., {\it Eur. Phys. J.} {\bf C7}
135, (1999).
%
\bibitem{Barbieri7} Y. Okada, K. Okumara and Y. Shimizu, {\it Phys. Rev.}
{\bf D61} 094001, (2000).
%
\bibitem{Chang} D. Chang, W. S. Hou and W. Y. Keung,
{\it Phys. Rev.} {\bf D48} 217, (1993).
%
\bibitem{Paradisi} P. Paradisi, {\it JHEP} {\bf 0602} 050, (2006).
%
\bibitem{MuLinYan} G. J. Ding, M. L. Yan,
{\it Phys. Rev.} {\bf D77} 014005, (2008).
%
\bibitem{AndiHektor} A. Hektor, Y. Kajiyama, K. Kannike,
hep-ph/0802.4015, (2008).
%
\bibitem{Rs1} L. Randall, R.Sundrum,
{\it Phys. Rev. Lett.} {\bf 83} 3370, (1999).
%
\bibitem{Rs2}L. Randall,
R.Sundrum, {\it Phys. Rev. Lett.} {\bf 83} 4690 (1999);
%
\bibitem{Goldberger}
W. D. Goldberger, M. B. Wise,  {\it Phys. Rev. Lett.} {\bf D 83}
4922, (1999).
%
\bibitem{Hisano} S. Chang, J. Hisano, H. Nakano, N. Okada, M.
Yamaguchi, {\it Phys. Rev.} {\bf D62} 084025, (2000).
%
\bibitem{Pamoral} A. Pomarol, {\it Phys. Lett.} {\bf B486} 153,
(2000).
%
\bibitem{Hewett} H. Davoudias, J. L. Hewett, T. G. Rizzo,
{\it Phys. Lett.} {\bf B473} 43, (2000).
%
\bibitem{Pamoral2} T. Gherghetta, A. Pomarol, {\it Nucl. Phys.} {\bf B586}
141, (2000).
%
\bibitem{Batell} B. Batell, T. Gherghetta,
{\it Phys. Rev.} {\bf D73}, 045016 (2006).
%
\bibitem{Huber4} S. J. Huber,C. A. Lee, Q. Shafi,  {\it Phys. Lett.} {\bf B531}
112, (2002).
%
\bibitem{Grossman} Y. Grossman, M. Neubert, {\it Phys. Lett.} {\bf B474}
361, (2000).
%
\bibitem{Huber} S. J. Huber, hep-ph/0211056, (2002).
%
\bibitem{Huber2} S. J. Huber, {\it Nucl. Phys.} {\bf B666} 269, (2003).
%
\bibitem{Huber3} S. J. Huber, Q. Shafi,
{\it Phys. Rev.} {\bf D63} 045010, (2001).
%
\bibitem{Kogan} I. I. Kogan,  S. Mouslopolous, A. Papazoglou, G. G. Ross,
{\it Nucl. Phys.} {\bf B615} 191, (2001).
%
\bibitem{Pamoral3} T. Gherghetta, A. Pomarol,
{\it Nucl. Phys.} {\bf B602} 3, (2001).
%
\bibitem{Ghoroku} K. Ghoroku, A. Nakamura,
{\it Phys. Rev.} {\bf D65} 084017, (2002).
%
\bibitem{KAgashe} K. Agashe, G. Perez and A. Soni,
{\it Phys. Rev.} {\bf D71} 016002, (2005).
%
\bibitem{EBlechman} K.Agashe, A. E. Blechman and F. Petriello,
{\it Phys. Rev.} {\bf D74} 053011, (2006).
%
\bibitem{Pree} E. D. Pree, M. Sher.,  {\it Phys. Rev.} {\bf D73},
 0955006 (2006).
 %
\bibitem{Moreau1} G. Moreau, J. I. S. Marcos, {\it JHEP} {\bf 0603}, 090
(2006)
%
\bibitem{Moreau2} F. Ledroit, G. Moreau, J. Morel, {\it JHEP} {\bf 09}
071 (2007).
%
\bibitem{MuckPilaftsisRuckl} A. Muck, A. Pilaftsis, and R. Ruckl,
{\it Phys. Rev.} {\bf D65}, 085037 (2002).
%
\bibitem{iltanZl1l2Extr} E. Iltan, {\it Eur. Phys. J.} {\bf C41},
233 (2005). 
\end{thebibliography}
\end{document}